\documentclass{emulateapj}

\newcommand{\hi}{\hbox{H {\sc i}}}
\newcommand{\hii}{\hbox{H {\sc ii}}}
\newcommand{\oii}{\hbox{[O {\sc ii}]}}

\newcommand{\kms}{\hbox{$\rm {km}~\rm s^{-1}$}}

\newcommand{\sini}{\hbox{${\rm sin}~i$}}
\newcommand{\sigoned}{\hbox{$\sigma_{1d}$}}
\newcommand{\logsigoned}{\hbox{${\rm log}~\sigma_{1d}$}}
\newcommand{\sigtwod}{\hbox{$\sigma_{2d}$}}
\newcommand{\vrot}{\hbox{$V_{rot}$}}

\newcommand{\wone}{\hbox{$S_{1.0}$}}
\newcommand{\whalf}{\hbox{$S_{0.5}$}}
\newcommand{\etal}{et al.\,}
\newcommand{\mydeg}{\hbox{$^\circ$}}
\newcommand{\taumod}{$\tau$-model}
\newcommand{\taumods}{$\tau$-models}

\long\def\hidetext#1{\relax}

\shortauthors{Weiner et al.}
\shorttitle{Survey of Galaxy Kinematics. II. Tully-Fisher Evolution}

\slugcomment{Accepted by ApJ}

\begin{document}

\title{A Survey of Galaxy Kinematics to $z \sim 1$ in the TKRS/GOODS-N 
Field. \\ II. Evolution in the Tully-Fisher Relation
\altaffilmark{1}}

\author{Benjamin J. Weiner\altaffilmark{2}, 
Christopher N.A. Willmer\altaffilmark{3,4}, 
S.M. Faber\altaffilmark{5}, 
Justin Harker\altaffilmark{5},
Susan A. Kassin\altaffilmark{5},
Andrew C. Phillips\altaffilmark{5},
Jason Melbourne\altaffilmark{5},
A.J. Metevier\altaffilmark{6,7}, 
N.P. Vogt\altaffilmark{8},
D.C. Koo\altaffilmark{5}
}

\altaffiltext{1}{Based in part on observations taken at the W.M. Keck
Observatory, which is operated jointly by the University of California
and the California Institute of Technology}
\altaffiltext{2}{Department of Astronomy, University
of Maryland, College Park, MD 20742, {\tt bjw@astro.umd.edu}.
Present address: Steward Observatory, University of Arizona, 
933 N. Cherry Av., Tucson, AZ 85721}
\altaffiltext{3}{Steward Observatory, University of Arizona, 
933 N. Cherry Av., Tucson, AZ 85721}
\altaffiltext{4}{On leave from Observatorio Nacional, Rio de Janeiro, Brasil}
\altaffiltext{5}{UCO/Lick Observatory, University of California, Santa Cruz, 
Santa Cruz, CA 95064}
\altaffiltext{6}{Center for Adaptive Optics, University of California, 
Santa Cruz, Santa Cruz, CA 95064}
\altaffiltext{7}{NSF Astronomy and Astrophysics Postdoctoral Fellow}
\altaffiltext{8}{Department of Astronomy, New Mexico State University,
P. O. Box 30001, Las Cruces, NM 88003}

\begin{abstract}

We use kinematic measurements of a large sample of galaxies from the
Team Keck Redshift Survey in the GOODS-N field to measure evolution in
the optical and near-IR Tully-Fisher relations to $z = 1.2$.  We
construct Tully-Fisher relations with integrated line-of-sight velocity
widths of $\sim 1000$ galaxies in $B$ and $\sim 670$ in $J$; these
relations have large scatter, and we derive a maximum-likelihood least
squares method for fitting in the presence of scatter.  The $B$-band
Tully-Fisher relations, from $z=0.4$ to $z=1.2$, show evolution of
$\sim 1.0-1.5$ mag internal to our sample without requiring
calibration to a local TF relation.  There is evolution in both
Tully-Fisher intercept and slope, suggesting differential luminosity
evolution.  In $J$-band, there is evolution in slope but little
evolution in overall luminosity.  The slope measurements imply that
bright, massive blue galaxies fade {\it more strongly} than fainter
blue galaxies from $z \sim 1.2$ to now.  This conclusion runs counter
to some previous measurements and to our naive expectations, but we
present a simple set of star formation histories to show that it
arises naturally if massive galaxies have shorter timescales of star
formation, forming most of their stars before $z\sim 1$, while less
massive galaxies form stars at more slowly declining rates.  This
model predicts that the higher global star formation rate at $z\sim 1$
is mostly due to higher SFR in massive galaxies.  The amount of fading
in $B$ constrains star formation timescale more strongly than redshift
of formation.  Tully-Fisher and color-magnitude relations can provide
global constraints on the luminosity evolution and star formation
history of blue galaxies.

\end{abstract}


\section{Introduction}

Surveys of large samples of galaxies at significant 
lookback times are a powerful instrument for measuring the
evolutionary history of galaxies.  First steps in these
probes include measuring number distributions of galaxies,
such as the galaxy luminosity function and color-magnitude
distribution.  A second step is the measurement of
scaling relations of galaxy properties.

Locally, relations between galaxy luminosity and
characteristic internal kinematic velocity -- the Tully-Fisher (TF)
relation for disks, the $D_n-\sigma$ relation for spheroidals
and its descendant, the Fundamental Plane -- are among the
tightest correlations and strongest tools for characterizing
galaxies.  These relations can be used to probe
evolution in galaxy properties, and simultaneously, a
successful scenario for galaxy formation and evolution 
will have to reproduce these relations, their scatter,
and their evolution with cosmic time.  However, measuring
galaxy internal kinematics at intermediate to high 
redshift is difficult since the sources are faint and 
high spectral resolution is needed.

In Paper I (Weiner \etal\ 2006) we presented 
measurements of galaxy kinematics from emission
lines, using Keck/DEIMOS spectra from 
the Team Keck Redshift Survey (TKRS) in the GOODS-N 
(Great Observatories Origins Deep Survey) field 
(Wirth \etal\ 2004; Giavalisco \etal\ 2004).  
The sample with kinematics from Paper I contains
$\sim 1000$ emission line velocity dispersions from integrated emission,
with median $<z> = 0.637$, and reaches beyond $z>1$.
We use integrated linewidths for fitting Tully-Fisher relations
in restframe $B$ and $J$ bands.  For a subsample of 380 galaxies,
we fit the 2-d spectra to model spatially resolved rotation and 
dispersion.  Paper I compared the 
properties of the spatially resolved rotation and dispersion
to validate the integrated linewidths.

A number of previous studies have measured luminosity-velocity 
relations from $0.1<z<1.0$.  Forbes \etal\ (1996),
Rix \etal\ (1997) and Mall{\' e}n-Ornelas \etal\ (1999)
used velocity dispersions from integrated emission for
$\sim 20$ galaxies each at median redshifts 0.48, 0.25 and 0.55.  
The pioneering
studies of Vogt \etal\ (1996, 1997) modeled rotation curves for
17 galaxies of disky morphology with median $<z> = 0.47$
by combining Keck/LRIS slitlet spectra with structural information from 
HST photometry.  Subsequent studies of rotation curves with similar
modeling procedures include Simard \& Pritchet (1998); Vogt (2000);
Ziegler \etal\ (2002) and Bohm \etal\ (2004);
Milvang-Jensen \etal\ (2003),
Bamford \etal\ (2005, 2006), and Nakamura \etal\ (2006);
Metevier \etal\ (2006);
and Conselice \etal\ (2005) who used $K$-band imaging of
galaxies in the HST Groth Strip.
These generally contained 20--100 galaxies with 
median redshifts $\sim 0.4 - 0.5$, and include field and
cluster samples.

These studies have produced a range of results for evolution in the
zeropoint of the Tully-Fisher relation; we discuss comparisons of TF
results in Section \ref{sec-othertf}.  TF measurements from resolved
rotation curves have mostly indicated a relatively mild amount of
magnitude evolution.  Previous samples are drawn from a variety of
populations and selection criteria.  In general, samples for measuring
rotation curves have been selected to be inclined disky objects with
measurable and reasonably orderly rotation.

By contrast,
the sample reported here is essentially selected only on magnitude and
emission line strength.  The lack of selection may be
important, given evidence that peculiar and disturbed galaxies are 
more common at high redshift (e.g. Abraham \etal\ 1996).
The TKRS sample is also
large and spans redshifts from 0.1 to 1.6, which allows
a measurement of Tully-Fisher relation evolution internal to the
sample rather than by comparison to a fiducial local Tully-Fisher
relation.  Our primary
tool for measuring kinematics is one-dimensional line-of-sight velocity
dispersion (linewidth), and secondarily the spatially resolved rotation 
profiles. Paper I discussed the properties of these velocity measures,
and showed that integrated linewidth as a measure of the
characteristic velocity of a galaxyis fairly robust against
observational effects and the details of galaxy kinematics,
and that we can measure velocities even for galaxies that are 
kinematically anomalous or not in orderly rotation, albeit 
with scatter due to the geometries of velocity fields and
inclinations.

We adopt an LCDM cosmology with $h=0.7$, $\Omega_M = 0.3$,
and $\Omega_\Lambda = 0.7$.  Magnitudes quoted in this paper are
in the AB system unless explicitly indicated as Vega.  
The sample and observations are discussed in Paper I;
Section \ref{sec-sample} briefly introduces the data.
Section \ref{sec-lwtf}
presents Tully-Fisher relations for integrated linewidths
and for rotation curves, and measures their evolution.
Section \ref{sec-othertf} compares to other TF measurements,
Section \ref{sec-evolcause} discusses the causes of evolution
in TF relation, and Section \ref{sec-models} uses luminosity 
evolution models to interpret the TF evolution.
In the Appendix we derive a maximum likelihood
method for fitting relations, such as our Tully-Fisher
relation, that have a substantial scatter.

\section{The data, Tully-Fisher samples, and kinematic properties}
\label{sec-sample}

\subsection{Photometric data and samples for Tully-Fisher}

The sample of galaxies used for study of the Tully-Fisher relation 
in this paper is drawn from the the Team Keck Redshift Survey (TKRS)
in the GOODS-N field (With \etal\ 2004).  The properties and completeness 
of the sample are described at length in Paper I; we summarize key
points here.

We construct restframe absolute magnitudes $M_B$ and $M_J$ 
and restframe colors $U-B$ and $R-J$ from 
SED fitting to the $BVRI$ and $HK^\prime$
magnitudes of Capak \etal\ (2004).  We use a
family of SEDs to determine the mapping from observed colors
to K-correction and restframe color at each galaxy's redshift.
The median errors on $M_B$ and $M_J$ are
0.11 and 0.14 mag respectively.
All magnitudes in this paper are on the AB system unless
otherwise noted.  

For the purpose of fitting Tully-Fisher relations, we exclude
the faint end of our galaxy sample.  The faintest galaxies
are only visible at low redshift, so are less useful for
comparisons across a range of redshifts.  Additionally,
the linearity of our observed Tully-Fisher relation can break down 
at faint magnitudes and small integrated linewidths.  We impose 
magnitude cuts of $M_B<-18$ and $M_J<-19$, which affect
the $z<0.5$ part of our sample.  Additionally, because
the spectroscopic sample is selected in apparent $R$,
the magnitude limit is a tilted line in the plane of restframe
$M_B$ and $U-B$, and the tilt evolves with redshift.
To test for any bias caused by the color-magnitude selection,
we also constructed a matched sample by imposing a 
single color tilt, and a magnitude limit that evolves
to track $L^*$ for blue galaxies; this limit is shown
as the diagonal lines in Figure 3
of Paper I, and
is discussed further in Section \ref{sec-fitsamples}.

\subsection{Kinematic measures and their properties}

Our primary kinematic measure is the line-of-sight dispersion of 
integrated emission \sigoned, measured 
in the TKRS DEIMOS spectra boxcar extracted to 1-D.
DEIMOS has a resolution of $\sigma_{inst}=1.4$ \AA\ with the 600 lines/mm 
grating as used for the TKRS, with 1.0\arcsec\ slits.  
For measuring \oii\ in a galaxy at $z=1$, the resolution corresponds
to $c \sigma_{inst} /\lambda_{obs} = 56$ \kms\ in the rest frame.
Although the resolution depends on $\lambda_{obs}$, because the wavelength 
range of the spectra is limited and different lines are used at
different redshifts, the resolution does not vary grossly with
redshift.  The observed widths of lines $\sigma_{obs}$ are measured by 
fitting Gaussians as described in Paper I, and the intrinsic line-of-sight 
velocity dispersion is computed from

\begin{equation}
\sigoned = \frac{c}{\lambda_{obs}} \sqrt{\sigma_{obs}^2 - \sigma_{inst}^2}.
\label{eqn-siginst}
\end{equation}

The emission line kinematic measure from the dispersion
of integrated emission is available for $\sim 90\%$ of galaxies
with redshifts on the blue side of the bimodal color distribution, 
with restframe $U-B<0.95$, but for very few red
galaxies, as discussed in Paper I.  For blue galaxies, the success 
rate for measuring linewidth is not strongly dependent on magnitude
or color.  However, for galaxies with low intrinsic dispersion 
$\sigoned \lesssim c \sigma_{inst}/\lambda_{obs}$,
it is difficult to measure \sigoned\ accurately.
As discussed in Paper I, it is possible for subtraction
in quadrature of $\sigma_{inst}$ to yield a formally very
small or negative intrinsic dispersion \sigoned, even though
it is not physically realistic.  Small \sigoned\ leads to
very large error on \logsigoned.  We refer to these galaxies,
with observed widths close to or less than instrumental,
as ``kinematically unresolved,'' and define them as meeting the criteria
on restframe dispersion
${\rm error}(\logsigoned)>0.25$, ${\rm error}(\sigoned)<30$,
and $\sigoned<25$ \kms.  

Rejecting these galaxies
would lead to a bias by preferentially rejecting low-velocity
galaxies, so for plotting and fitting purposes we assign them
a low velocity, $\logsigoned = 1.4 \pm 0.2$  ($\sigoned = 25$ \kms).
The results of fitting do not depend strongly on the exact value assigned.
We do reject galaxies with $\sigoned>25$ \kms\ and 
${\rm error}(\logsigoned)>0.25$; for these, the large error is 
usually a sign of a bad fit or data contaminated by night sky lines.
Including kinematically unresolved galaxies, there are 913
galaxies with acceptable \sigoned\ and $M_B<-18$, and 647 with \sigoned\ 
and $M_J<-19$.  In Section \ref{sec-fitmethods} we outline a fitting 
method which treats these unresolved galaxies more robustly by fitting 
the ensemble of observed width $\sigma_{obs}$ before the 
instrumental resolution is subtracted. 

For a subset of galaxies selected on size and emission line 
strength, we also fit a kinematic model to the 2-d spectrum, 
measuring the spatially resolved kinematics with line-of-sight 
terminal velocity of the rotation
curve \vrot, and dispersion \sigtwod.  380 galaxies yielded
good spatially resolved fits.  The dispersion 
\sigtwod\ from the spatially resolved fits differs from the
dispersion computed from integrated linewidth \sigoned\
because the integrated linewidth also includes the velocity
spread caused by the rotation gradient.

The modeling of velocity fields in Section 4 of Paper I shows 
how the integrated linewidth can represent the velocity spread
in the full galaxy velocity field, because the linewidth
arises from different velocities mixed together by seeing.
In Section 5 of Paper I we also compared the integrated kinematic 
measure \sigoned\ with the measures from resolved fits, \vrot\ and
\sigtwod, and with combinations of the resolved measures,
e.g. $\whalf^2 = 0.5 \vrot^2 + \sigtwod^2$.  We
concluded that \sigoned\ correlates well with the
combination of the resolved measures, and as a
probe of galaxy internal kinematics, is fairly robust against
observational effects such as slit position angle, although there
is scatter in \sigoned\ due to galaxy properties such as 
inclination.  We found
that galaxies with spatially resolved kinematics fall
on a range between rotation dominated, $\vrot/\sigtwod>1$,
and dispersion dominated, $\vrot/\sigtwod<1$;
the 1-d linewidth \sigoned\ and the combined velocity
scale \whalf\ are better proxies for kinematics than
either \vrot\ or \sigtwod\ alone.  

We found that a
significant fraction of $z \sim 1$ elongated galaxies 
do not show strong rotation, implying that ellipticity 
is not always a good measure of inclination, or that some
galaxies are not rotating disks, or both.  
Because the conventional assumptions about inclined rotating
disks may break down, using galaxy ellipticity to perform
inclination corrections on a sample of integrated linewidths
\sigoned\ is not necessarily justified.  However, any
unrestricted sample of galaxies does include a range of
inclinations, so the mean value of \sigoned\ is reduced
by some factor when inclination corrections are not applied.

\section{The Tully-Fisher relation for integrated linewidths}
\label{sec-lwtf}

\subsection{Tully-Fisher relations in redshift subranges}
\label{sec-tfzbins}

\begin{figure*}[ht]
\begin{center}
\includegraphics[width=5.5truein]{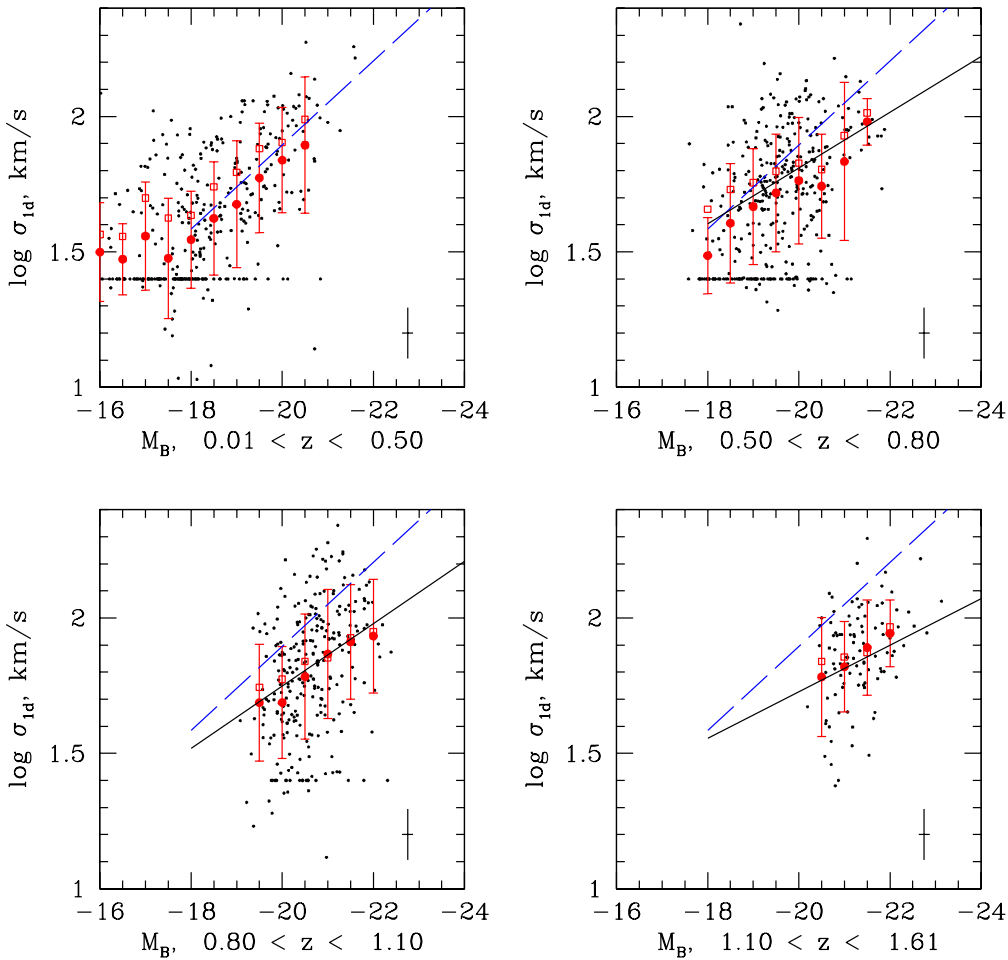}
\caption{The Tully-Fisher relation in the TKRS for integrated 
line-of-sight linewidth and rest $B$ magnitude, in four
redshift ranges.  Individual galaxies are plotted as small
points.  Large circles and errorbars are the unweighted mean
and RMS in magnitude bins; large squares are the weighted mean.  
The small points at $\logsigoned=1.4$
are kinematically unresolved galaxies.
The dashed diagonal line is
a fit to the low-redshift range for $M_B<-18$, and is
repeated in all four panels.
The solid diagonal lines are fits to the higher redshift ranges.
The cross in the lower right corner of the plots shows the
median observational errors.  At high redshifts, nearly all the 
galaxies fall below or to the right of the low-redshift relation.
}
\label{fig-lwtf}
\end{center}
\end{figure*}

\begin{figure*}[ht]
\begin{center}
\includegraphics[width=5.5truein]{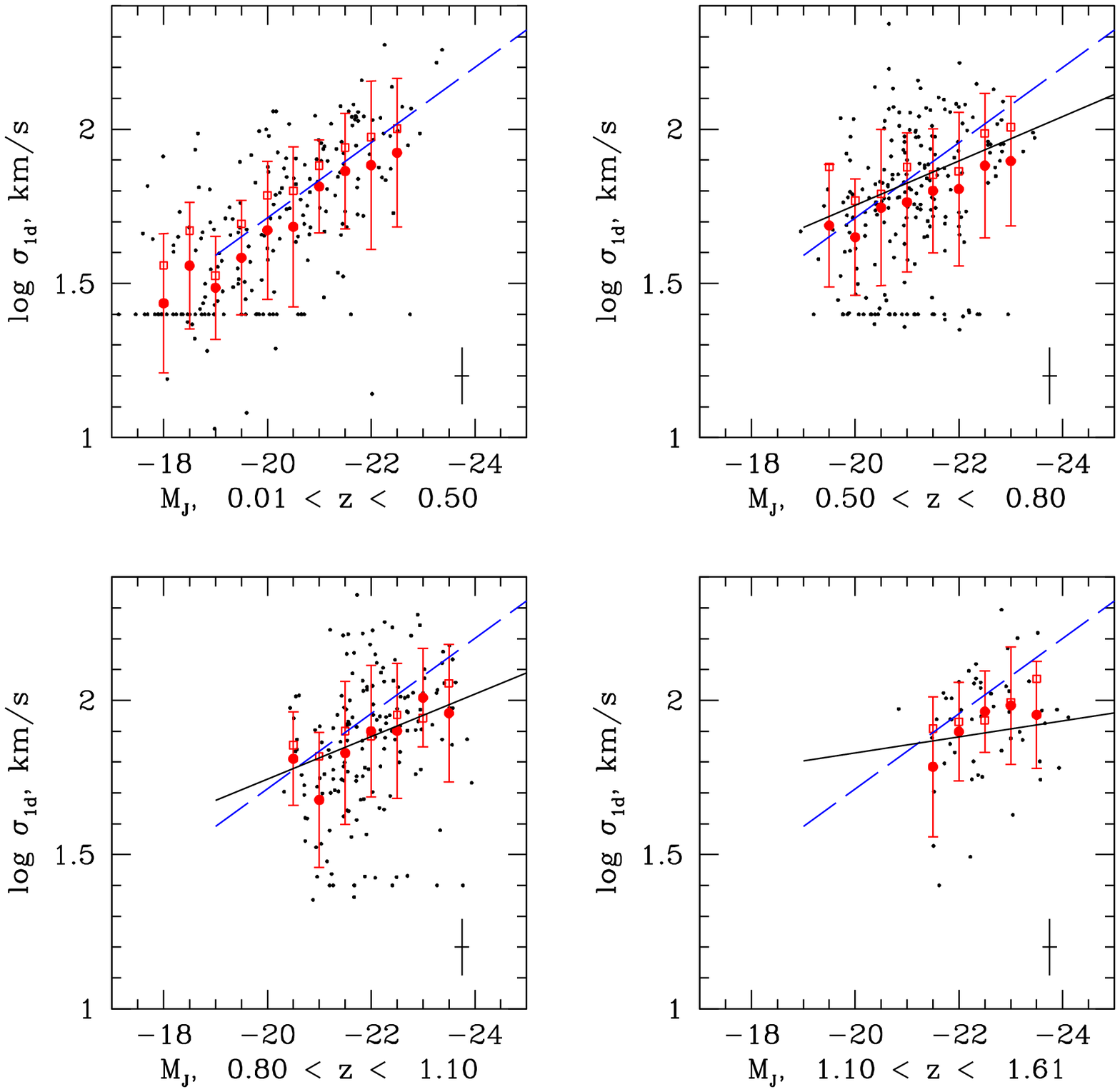}
\caption{The Tully-Fisher relation in the TKRS for integrated 
line-of-sight linewidth and rest $J$ magnitude, in four
redshift ranges.  Individual galaxies are plotted as small
points.  Large circles and errorbars are the unweighted mean
and RMS in magnitude bins; large squares are the weighted mean.
The small points at $\logsigoned=1.4$
are kinematically unresolved galaxies.
The dashed diagonal line is
a fit to the low-redshift range for $M_J<-19$, and is
repeated in all four panels.
The solid diagonal lines are fits to the higher redshift ranges.
The cross in the lower right corner of the plots shows the
median observational errors.  
The $J$-band relation shows less evolution than in the $B$-band.
}
\label{fig-jlwtf}
\end{center}
\end{figure*}

We combine the absolute magnitudes and the line-of-sight linewidths
of integrated emission \sigoned\ to produce the Tully-Fisher
relations for integrated linewidths shown in Figure \ref{fig-lwtf}
for $B$-band and Figure \ref{fig-jlwtf} for $J$-band.
The sample is broken into several redshift ranges.
In each figure,
the large points show weighted and unweighted means of
\logsigoned\ in bins of magnitude.  The
diagonal lines are maximum likelihood fits, described below,
to the observed linewidths $\sigma_{obs}$ in each 
redshift range; we fit to $\sigma_{obs}$  to mitigate numerical
issues caused by the subtraction of instrumental resolution,
discussed below.  The low redshift fit (dashed line) is
plotted in all four redshift ranges for comparison.  
The fits are restricted to galaxies brighter than $M_B=-18$
and $M_J=-19$ respectively.  Below these magnitude limits,
\sigoned\ becomes small, comparable to random non-gravitational
motions, and difficult to measure, and the linearity of the 
observed Tully-Fisher relation in \sigoned\ is dubious.

There is clear evolution in the $B$-band Tully-Fisher relation: 
at high redshift,
galaxies lie to the right of or below the low-$z$ line
(brighter or lower velocity).  We discuss fitting
methods and interpretations of the evolution below.

In these and all subsequent figures and discussion of the 
Tully-Fisher relation, we plot, fit and discuss velocity as a function
of magnitude, $V(M)$.  For these high-redshift galaxies, the
primary selection boundary is the magnitude limit, and
the errors are larger on velocity than on magnitude.
For local galaxies, the convention is often
to treat luminosity as a dependent variable, in part due to
the history of the TF relation as a distance indicator.
Taking velocity as the dependent variable
is often called the ``inverse Tully-Fisher relation'' (Fouque \etal\ 1990).
We always perform fits that take both errors into account, so 
there is not an independent/dependent distinction in the fit, but
the magnitude selection introduces an asymmetry that makes fitting
$V(M)$ more sensible.
The reader should keep in mind that the slope of the fit
lines in Figures \ref{fig-lwtf} and \ref{fig-jlwtf}
is opposite the sense in which $M(V)$
Tully-Fisher relations are usually described as shallow or steep.

Figure \ref{fig-lwtf} plots the galaxies in the ``good linewidth'' 
sample individually as small points.  
The large points in Figure \ref{fig-lwtf} show the weighted 
and unweighted means of \logsigoned\ in magnitude bins; the
error bars indicate the RMS within each bin (not the error
of the mean).  Large points are only plotted for bins with
$\geq 5$ galaxies. 
The mean-in-bins points show that there is a clear relation
of velocity with magnitude within each redshift range, and that
in $B$, this relation evolves with redshift.

There is a significant scatter
induced by intrinsic scatter in the TF relation, and secondarily by
errors on the individual velocity measurements.  
For galaxies with $M_B<-18$, the intrinsic scatter about the
fitted relations is 0.18 dex 
in \logsigoned, and the median observational error is 0.084 dex.
Intrinsic scatter in the \logsigoned\ TF relation could come from
scatter in true properties of the galaxies, from the transformation
from detailed galaxy kinematics to integrated linewidth, and from 
the lack of inclination correction.  In Section 4.3 of Paper I we 
showed that for idealized circular rotating disks, omitting inclination
and extinction corrections leads to scatter of $\sim 0.19$ dex in
a $B$-band TF.  

However, the similarity between the idealized and 
measured scatter is fortuitous.  When we use the ellipticities
from HST/ACS imaging to infer inclination and extinction 
corrections for the $M_B<-18$ sample, we find that the intrinsic
scatter about fitted TF relations 
{\it does not decrease}.\footnote{For the fits comparing the 
effect of inclination and extinction corrections, we exclude
galaxies with ellipticity $e<0.2$, to avoid large corrections to 
velocity width.}  The scatter increases minutely from 0.18
to 0.19 dex.  The likely cause is that there are a significant
number of galaxies that are not consistent with circular
rotating disks, as discussed in Paper I.  For these galaxies,
inclination corrections to velocity may be inappropriate
and/or ellipticities may not give proper inclinations.
We discuss TF scatter further in Section \ref{sec-othertf}.
Because it is not clear that the ``corrections'' are an
improvement for this sample, which is not restricted to
disky morphologies, we refrain from applying them.

\subsection{Fitting the Tully-Fisher relation: methods and sample}
\label{sec-fitmethods}

In each redshift range, we fit linear Tully-Fisher relations
with the ridgeline

\begin{equation}
\logsigoned = A_\lambda + B_\lambda (M_\lambda - M_{zp,\lambda}),
\label{eqn-tfrel}
\end{equation}

\noindent
with intrinsic scatter $C_\lambda$ in \logsigoned.
The TF relations are zeropointed at $M_{zp,B}=-21$ and $M_{zp,J}=-22$ 
to decrease covariance between intercept $A$ and slope $B$. 
We used a maximum likelihood method (hereafter MLS, for maximum
likelihood with scatter) to fit the Tully-Fisher
relation to data that have intrinsic scatter and errors in both 
coordinates.  The MLS method 
is derived in the Appendix,
and tests of fitting methods are discussed in Section \ref{sec-fittest}.

The MLS method treats the TF relation with scatter as a model
probability distribution, and convolves the model with the
error distributions of the observations to compute a conditional
probability of the model given the data.
For a linear ridgeline and gaussian scatter and errors, 
the MLS method becomes mathematically very similar to a generalized 
least squares (GLS) method, based on the {\tt fitexy} routine 
(Press \etal\ 1992).  GLS performs a least-squares fit by 
adding the intrinsic scatter to
the error in the $y$-coordinate, here taken to be the velocity
(see Tremaine \etal\ 2002; Novak, Faber \& Dekel 2005; Pizagno \etal\ 2005;
Bamford \etal\ 2006).
The MLS method places this {\it ad hoc} extension on a firmer
statistical footing, and yields nearly identical fit results.

\begin{figure}[ht]
\begin{center}
\includegraphics[width=3.5truein]{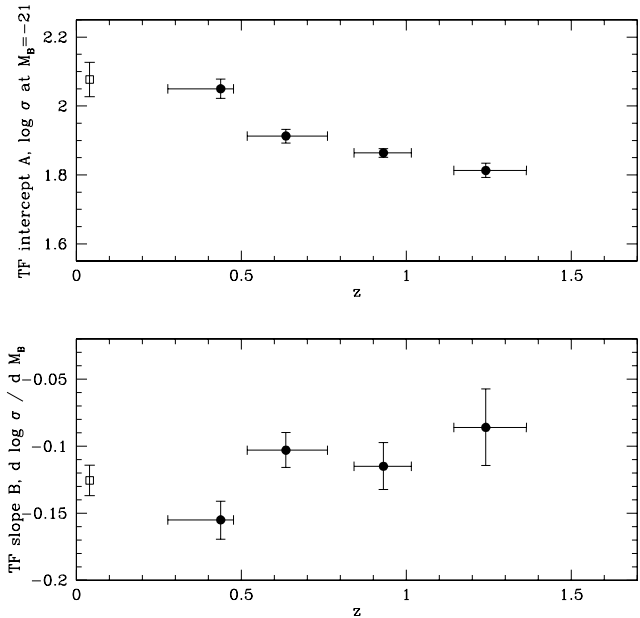}
\caption{Redshift evolution of the $B$-band Tully-Fisher intercept
and slope, $\logsigoned = A_B + B_B (M_B+21)$.  The intercept
and slope are fitted with the MLS maximum likelihood method 
as described in Section \ref{sec-fitmethods} in four redshift ranges.  
The points are plotted at the median redshift of the range
and the horizontal bars are the 68\% range in redshift. 
The local $B$-band TF intercept and slope of Sakai \etal\ (2000)
are plotted at low $z$ for comparison.
}
\label{fig-paramevol}
\end{center}
\end{figure}

\begin{figure}[ht]
\begin{center}
\includegraphics[width=3.5truein]{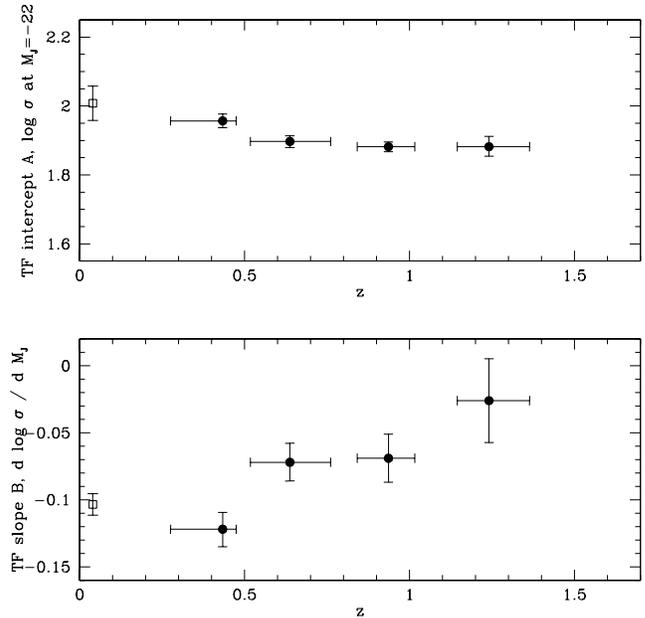}
\caption{Redshift evolution of the $J$-band Tully-Fisher intercept
and slope, $\logsigoned = A_J + B_J (M_J+22)$.  The intercept
and slope are fitted with the MLS maximum likelihood method 
as described in Section \ref{sec-fitmethods} in four redshift ranges.  
The points are plotted at the median redshift of the range
and the horizontal bars are the 68\% range in redshift. 
The local $J$-band TF intercept and slope of Watanabe \etal\ (2001)
are plotted at low $z$ for comparison.
}
\label{fig-jparamevol}
\end{center}
\end{figure}

Because some of the galaxies are kinematically unresolved,
we perform the MLS fits by convolving the model distribution with
the data in the space of observed 
linewidth $\sigma_{obs}$ rather than intrinsic dispersion $\sigoned$.
This enlarges the samples with good measurements slightly, 
to 968 galaxies brighter than $M_B=-18$, and 677 brighter than
$M_J=-19$, each with a $\sigma_{obs}$
measurement from its strongest emission line.  
The best-fit relations in each redshift range from these fits 
are shown in Figures \ref{fig-lwtf} and \ref{fig-jlwtf}.
The best-fit intercept and slope are listed in Table 
\ref{table-bestpars} and plotted against redshift in Figures
\ref{fig-paramevol} and \ref{fig-jparamevol}.
The best-fit values from fitting directly in $\sigoned$ space 
are statistically similar, as long as the kinematically unresolved objects
are included by setting them to the arbitrary value of $\logsigoned=1.4$.

The MLS method can fit $A,B,$ and $C$ together; here we fixed
the intrinsic scatter at $C=0.18$ dex, which is the best value for the
entire sample in both $B$ and $J$.  The TF slope and intercept are only 
very weakly dependent on the scatter, as long as the assumed scatter is 
reasonable (within a factor of 1.5--2).  Because the scatter is 
sensitive to the error estimates, and there are a significant number
of kinematically unresolved galaxies, and the sample is not 
extremely large, we do not trust it to measure evolution in the
scatter.

\subsubsection{Properties of the data set}
\label{sec-dataprops}

There has been some argument over proper methods for fitting
Tully-Fisher and similar relations (see Willick 1994), and a related 
controversy has flared briefly in the field of black hole
masses (see Tremaine \etal\ 2002; Novak \etal\ 2005).  It is not
our intention to present a solution for fitting model relations 
to data that works in all cases, but to present a well-posed model 
that can be constrained fairly and applied to our dataset.  
We list here some salient features of high-redshift 
Tully-Fisher relations and our dataset in particular.
Section \ref{sec-fittest} tests the reliability of fitting 
methods using simulated data that reproduces these properties.

1) The data available in high-$z$ TF relations are strongly 
selected in magnitude.
A sample is unlikely to cover more than $\sim 3$ magnitudes at
any given redshift: the faint limit is set by S/N and the bright
limit is set by the lack of galaxies brighter than $M^*$, and the
magnitude range of a sample narrows at higher redshift.  However,
there is not a strong selection on velocity.
The selection limits make it more straightforward to use the 
``inverse'' TF relation in which magnitude is the independent 
variable.

2) The (inverse) Tully-Fisher relation has a shallow slope.
Typical slopes of log velocity on magnitude are 0.125 to 0.09
from $B$ to $K$;
log~$V =$ (0.31 to 0.23) log~$L + const$
(e.g. Tully \etal\ 1998; Sakai \etal\ 2000).

3) At high $z$, observational errors on log velocity or dispersion 
are larger than errors on magnitude, when the slope is taken 
into account.  Our median errors are 0.084 dex in \logsigoned,
0.12 mag in $M_B$, and 0.14 mag in $M_J$, but since the slope is 
$0.08-0.15$ dex/mag, the errors on \logsigoned\ are effectively $4-9$ 
times larger than on magnitude.  Fortunately, the large velocity
errors scatter along the magnitude selection limit rather
than across it.

4) The intrinsic scatter in the high-$z$ Tully-Fisher relation is 
large.  This is especially true for a sample like ours which is 
not highly selected on morphology, kinematic regularity, or other 
properties, and also not corrected
for inclination.  Fitting methods which do not take into 
account intrinsic scatter are generally biased, especially in slope.

5) The TKRS resolution of 1.4 \AA\ is only moderately high, so
some low-linewidth objects are kinematically unresolved, and others
which are barely resolved have large fractional error on \sigoned.
Therefore the error on \sigoned, and even more so on \logsigoned,
is inversely correlated with \sigoned.  This can lead to biases
in fitted relations and weighted means: the weighted mean of 
\logsigoned\ will be biased high if gaussian errors on \logsigoned\
are assumed.  Fitting in wavelength
space $\sigma_{obs}$ partially mitigates this bias.

A correlation between \logsigoned\ and error on \logsigoned\
shows up as an offset between weighted and unweighted means of
\logsigoned.  The binned points in Figures \ref{fig-lwtf} and 
\ref{fig-jlwtf} show that this offset exists in the $0<z<0.5$ 
data but is relatively small in the higher redshift data.
Because we use a fit method that accounts for both the errors
on individual points and the intrinsic scatter, the best fit line
lies between the weighted and unweighted means.

6) The raw observational errors on velocity and magnitude 
are independent.  However, transforming to deprojected
properties can make the errors covariant; inclination 
and extinction corrections are highly correlated.

\subsubsection{Magnitude-limited samples}
\label{sec-fitsamples}

We only include galaxies brighter than $M_B<-18$ or $M_J<-19$
in the Tully-Fisher fits,
for two reasons.  First, very low luminosity galaxies are not visible
at higher redshift and we want to keep the samples comparable. Second, 
the low luminosity galaxies have very low dispersion, close to
the random velocities in \hii\ regions and at the limit of
what we can resolve, so there is effectively a minimum linewidth.
Faint galaxies tend to scatter up in dispersion and would
flatten the TF relation slope if included in the fits; this 
effect is seen in the low-redshift panel
of Figure \ref{fig-lwtf}.

Because we are fitting TF relations over a wide redshift range,
the relative depth of the sample varies as a function of $z$.
Since the sample is selected in observed $R$ band, the 
changes in $K$-correction with redshift
also change the slope of the magnitude limit with restframe color,
as shown in Figure 3
in Paper I.  Redder
galaxies are progressively more disfavored at higher redshift.
To match samples of similar galaxies at different redshifts, 
we constructed a near volume-limited sample.  This sample meets a 
rolling magnitude cut that corresponds to the limiting color-magnitude
line at $z = 0.9$, and tracks the luminosity evolution of the
blue galaxy population by evolving by 1.3 magnitudes per unit
redshift, as does $L^*$ (Willmer \etal\ 2006).  
The magnitude limit has a constant slope with color, to guard 
against spurious trends that might be induced by favoring
bluer galaxies at high redshift.  The restricted sample
is approximately volume limited with respect to mass for $z<0.9$,
if we assume that the primary evolution is fading in luminosity.
This assumption is not proven, but the blue LF evolution and 
our discussion of models in Section \ref{sec-taumodels} provide
some evidence for it.

The equation of
the cut is $M_B < -18.5 - 2*(U-B) + 1.3*(0.9-z)$, and it is shown
as the diagonal lines in Figure 3
of Paper I.  We also restricted the sample to galaxies on the blue
side of the color bimodality, with $U-B<0.95$.
659 galaxies meet these cuts and have acceptable \logsigoned\ 
meeting the error cuts described previously; of these
49 are kinematically unresolved.  For the fits done in wavelength
space $\sigma_{obs}$,  672 galaxies meet the cuts.
The best fit $B$-band TF parameters for the $M_B<-18$ and 
the rolling magnitude limit
samples are listed in Table \ref{table-bestpars}.  The differences
between the samples are minor at most; only the difference
in the intercept $A$ in the lowest redshift range is marginally
statistically significant.


When we split the sample into aligned and misaligned samples based on
slit alignment with the galaxy major axis, the fits are not
significantly different within the errors, consistent with the lack of
dependence of \logsigoned\ on slit alignment shown in Section 4.4
of Paper I.

\subsection{Results: redshift evolution in the Tully-Fisher relation}
\label{sec-tfevol}

Figures \ref{fig-lwtf} and \ref{fig-jlwtf} show the
redshift evolution of the Tully-Fisher relation for 
integrated linewidths.  
It is clear that in $B$-band, either galaxy properties
or the relation of observables to properties are
evolving, since almost all the points in the last redshift range
lie below or to the right of the low-redshift relation.
Here the size of our sample is important, since we can
measure evolution within the sample and do not have to rely
solely on calibrating the intercept to a local relation.

\begin{figure}[ht]
\begin{center}
\includegraphics[width=3.5truein]{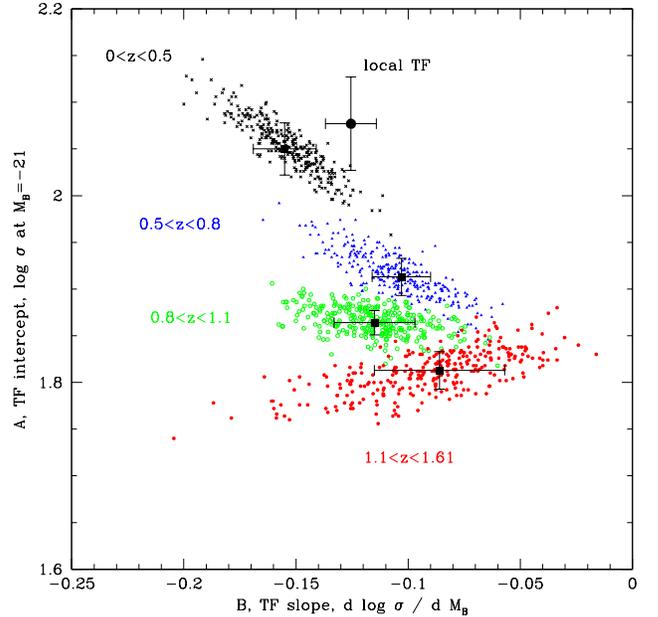}
\caption{Error estimation on $B$-band
Tully-Fisher parameters slope $B$ versus intercept $A$ in
four redshift ranges.
Each point represents a fit to one of 300 bootstrap
resamplings of the data in that $z$ range, and the locus of
points indicates the covariance and scatter in the parameters;
Xes are $0<z<0.5$, triangles $0.5<z<0.8$, open circles $0.8<z<1.1$,
and filled circles $1.1<z<1.61$.  The filled squares and error bars
near the centers of the bootstrap distributions show the best-fit
values from Table \ref{table-bestpars}.  The local $B$-band 
intercept (converted to \sigoned) and slope are shown as the large
point (Sakai \etal\ 2000).  Intercept and slope are quite covariant 
in the $0<z<0.5$ sample.  There is clear evolution in intercept with
redshift, and some evidence for evolution in slope.}
\label{fig-parboot}
\end{center}
\end{figure}

\begin{figure}[ht]
\begin{center}
\includegraphics[width=3.5truein]{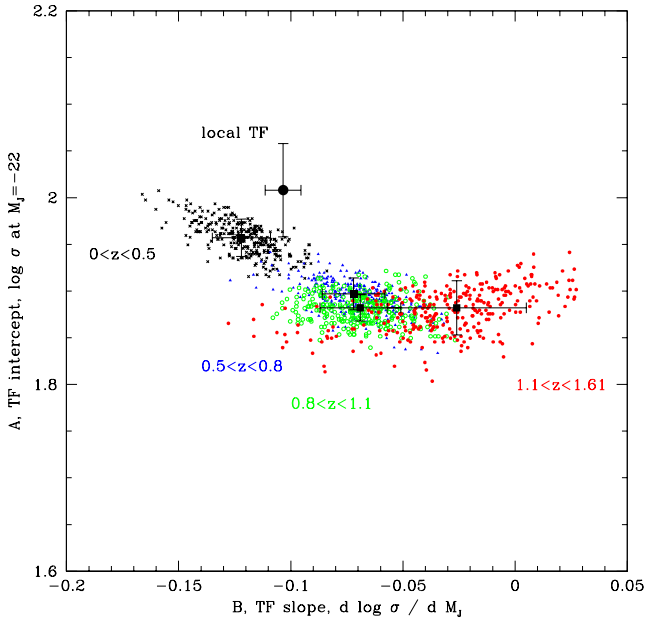}
\caption{Error estimation on $J$-band
Tully-Fisher parameters slope $B$ versus intercept $A$ in
four redshift ranges.
Each point represents a fit to one of 300 bootstrap
resamplings of the data in that $z$ range, and the locus of
points indicates the covariance and scatter in the parameters;
Xes are $0<z<0.5$, triangles $0.5<z<0.8$, open circles $0.8<z<1.1$,
and filled circles $1.1<z<1.6$.  The filled squares and error bars
near the centers of the bootstrap distributions show the best-fit
values from Table \ref{table-bestpars}.
The local $J$-band intercept (converted to \sigoned)
and slope are shown as the large point (Watanabe \etal\ 2001).
Evolution in intercept is weak, but there is
evidence for evolution in slope.}
\label{fig-jparboot}
\end{center}
\end{figure}

At higher redshifts, the galaxies fall below or to the right
of the low-redshift fit line in $B$, meaning they have lower velocities
or are brighter.  In $J$-band, the offset between high-redshift points
and low-redshift fit is smaller than in $B$.
The remainder of this paper is devoted to
quantifying the evolution in intercepts and slope, determining whether 
it is a change in the basic physical properties of galaxies or only 
a change in the kinematic tracers, and presenting a few toy models 
for evolution in the TF relation.

The upper panels of Figures \ref{fig-paramevol} and  
\ref{fig-jparamevol} show the 
evolution in TF intercept $A$ as a function of redshift in
$B$ and $J$-bands,
and the intercept of local TF relations for comparison 
(Sakai \etal\ 2000; Watanabe \etal\ 2001).  
Here and afterwards we show results from the full samples 
with $M_B<-18$ and $M_J<-19$.  

There is substantial evolution in the $B$-band
intercept, internal to our sample and compared to the local
sample.  However, there is
relatively little evolution in the $J$-band intercept.
These agree with the visual impression of Figure \ref{fig-lwtf},
where most of the high-redshift galaxies lie below or to
the right (brighter $M_B$) of the low-redshift fit, while in 
Figure \ref{fig-jlwtf} the high-redshift galaxies are not very
offset from the low-redshift fit.

For the local TF relations, we transform the relations
from inclination-corrected \hi\ width $W_{20,corr}$
to the intercept $A$ in \logsigoned.  We assume $\sigoned = 0.28 W_{20}$
(Kobulnicky \& Gebhardt 2000), and decorrect the local TF relations 
by the median inclination and extinction corrections for a randomly oriented
sample, using extinctions $A_{Bext} = 1.57~{\rm log}(a/b)$ 
and $A_{Jext} = 0.736~{\rm log}(a/b)$, as used in the local TF relations, but
omitting velocity-dependence of extinction.\footnote{Extinction 
corrections that depend on velocity tend to change the $B$-band slope
by about 10\% and the $J$-band slope relatively little.  We avoid
using any velocity-dependent extinction for the linewidth TF relation
because our velocity measure is noisy and it could induce spurious
correlations.}  The median inclination for a randomly oriented sample
is 60\mydeg, so that the median $<{\rm log}~\sini> = -0.0625$ and 
$<{\rm log}(a/b)=0.30$.
We assigned a systematic error of 0.05 dex to this conversion 
from $W_{20,corr}$ to $A$; error in the conversion
dominates over the statistical errors in the local fits.
Other local $B$-band TF relations lie in essentially the same place 
as the Sakai \etal\ relation (e.g. Tully \& Pierce 2000).

The lower panels of Figures \ref{fig-paramevol} and \ref{fig-jparamevol} 
plot the trend of slope $B_\lambda$ with redshift, and show some
evidence for TF slope evolution, seen in the fits plotted in
Figures \ref{fig-lwtf} and \ref{fig-jlwtf}.  
The $z>0.5$ redshift points all have shallower inverse-TF slopes
than both the low redshift points and the local $B$ and $J$-band 
TF relations.  In the $B$-band, the total significance
of the high-$z$ slope offsets from our $0<z<0.5$ value is 
3.0 sigma.\footnote{A similar slope evolution is found in the
forthcoming larger DEEP2 sample, which will provide a more
precise measurement.}
The $J$-band shows a similar trend, though there are too few
galaxies in the $J$-band sample at $z>1.1$ to measure
a reliable slope.

Figures \ref{fig-lwtf} and   
\ref{fig-paramevol},
and Table \ref{table-bestpars} show that there is evolution in
both the intercept and slope of the $B$-band linewidth Tully-Fisher
relation.  In the lowest redshift range, the slope is $B=-0.155 \pm 0.014$,
or $-1/B= 6.5 \pm 0.6$, at most modestly different from 
local measurements of the $B$-band
Tully-Fisher slope, e.g. $-1/B=7.3$ (Tully \& Pierce 2000) 
or $8.0 \pm 0.7$ (Sakai \etal\ 2000).  Given the effects on
slope of inclination-induced scatter (Section 4.3
of Paper I), velocity-dependent extinction,
and the covariance between $A$ and $B$ shown in Figure \ref{fig-parboot},
the intercept and slope in our lowest redshift range are not
significantly different from the local values.

The $B$-band TF evolution appears
to be strongest from low redshift to $z \sim 1$, 
but covariance of slope and intercept plays 
a role here, as shown in Figure \ref{fig-parboot}.  It is 
possible that our $0<z<0.5$ range has an intercept $A$ that is
a bit too high and a slope $B$ that is a bit too steep, compared
to the local values.  This could arise in part from the correlation
of error(\sigoned) and \sigoned, point 5 of Section 
\ref{sec-dataprops}.

Because the fit parameters $A$ and $B$ are covariant,
we show confidence regions generated
by bootstrap resampling in Figures \ref{fig-parboot} and \ref{fig-jparboot}, 
to give a more complete picture
of the evolution of TF parameters.  For each of the four
redshift ranges, we generated 300 samples with replacement
from the original data and refit these samples.  The loci of points 
indicate the error ranges and covariance.  Intercept evolution
is very significant in $B$-band, but only significant in $J$ if
the local sample is used as the calibrator.
The fits suggest, at $\sim 3\sigma$, that the high-redshift slope 
$B_\lambda$ is shallower in velocity on magnitude $V(M)$ in both bands; 
they provide a strong rejection of the
idea that the high-redshift slope is steeper in $V(M)$.

Galaxies at higher redshift have lower \logsigoned\ at a given 
magnitude, or equivalently a brighter magnitude at a fixed dispersion.
The luminosity evolution is physically more likely than
velocity evolution, assuming that we are seeing evolution 
in the global properties of galaxies rather than just in
the properties of \sigoned\ as a kinematic tracer, as discussed
further in Section \ref{sec-evolcause}.
Stellar populations are well-known to evolve in luminosity,
while the characteristic density and velocity of the inner part of 
a galactic halo change little at late stages of its mass accretion 
history, because late-time accreted mass is
relatively low density (Wechsler \etal\ 2002).
The fact that evolution is stronger in the $B$-band
than in the $J$-band supports the interpretation of
luminosity evolution.

Since there appears to be slope evolution, the amount of
magnitude evolution is itself a function of magnitude.
In the highest redshift range, with median $z=1.2$,
a galaxy with $M_B=-22$ must fade by $\sim 2$ mag to reach
the low-redshift $z=0.4$ TF relation, while a galaxy with $M_B=-20$ 
must fade by only 1.2 mag.  These numbers will be reduced by $\sim 0.5$ mag
if the $z=0.4$ relation is biased high in intercept as discussed above.
We discuss models for differential
luminosity evolution in Section \ref{sec-taumodels}.

\begin{deluxetable*}{llrllrrr}

\tablecaption{
Best-fit Tully-Fisher relations
\label{table-bestpars}
}

\tablecolumns{8}
\tablewidth{0pt}
\tabletypesize{\small }

\tablehead{
Band & Sample & Number & Redshift & Median   & Zeropoint &
  \multicolumn{2}{c}{Maximum likelihood fit\tablenotemark{a}} \\
    &     &     & range  & redshift & (mag) & intercept $A_\lambda$ & slope $B_\lambda$
}
 \startdata

$B$ & $M_B<-18$ &  218   & $0.07<z<0.5$ & 0.437 & --21 &
 $2.050 \pm 0.028$ & $-0.155 \pm 0.014$ \\
$B$ & $M_B<-18$ &  374   & $0.5<z<0.8$  & 0.635 & --21 &
 $1.913 \pm 0.020$ & $-0.103 \pm 0.013$ \\
$B$ & $M_B<-18$ &  280   & $0.8<z<1.1$  & 0.931 & --21 &
 $1.864 \pm 0.013$ & $-0.115 \pm 0.018$ \\
$B$ & $M_B<-18$ &   96   & $1.1<z<1.61$ & 1.241 & --21 &
 $1.813 \pm 0.020$ & $-0.086 \pm 0.029$ \\
\tableline
$B$ & Blue + magcut\tablenotemark{b} & 87 & $0.11<z<0.5$ & 0.438 & --21 &
 $2.007 \pm 0.037$ & $-0.154 \pm 0.028$ \\
$B$ & Blue + magcut & 237 & $0.5<z<0.8$  & 0.679 & --21 &
 $1.923 \pm 0.024$ & $-0.119 \pm 0.020$ \\
$B$ & Blue + magcut & 252 & $0.8<z<1.1$  & 0.927 & --21 &
 $1.864 \pm 0.013$ & $-0.122 \pm 0.018$ \\
$B$ & Blue + magcut &  96 & $1.1<z<1.61$ & 1.241 & --21 &
 $1.813 \pm 0.020$ & $-0.086 \pm 0.029$ \\
\tableline
$J$ & $M_J<-19$ &  165   & $0.09<z<0.5$ & 0.433 & --22 &
 $1.957 \pm 0.020$ & $-0.122 \pm 0.013$ \\
$J$ & $M_J<-19$ &  232   & $0.5<z<0.8$  & 0.637 & --22 &
 $1.897 \pm 0.017$ & $-0.072 \pm 0.014$ \\
$J$ & $M_J<-19$ &  165   & $0.8<z<1.1$  & 0.936 & --22 &
 $1.882 \pm 0.014$ & $-0.069 \pm 0.018$ \\
$J$ & $M_J<-19$ &   50   & $1.1<z<1.53$ & 1.241 & --22 &
 $1.882 \pm 0.029$ & $-0.026 \pm 0.031$ \\
 \enddata

\tablenotetext{a}{Maximum likelihood fits with TF scatter of 0.18 in \logsigoned.}
\tablenotetext{b}{Blue+magcut sample restricted to $U-B<0.95$, and color-magnitude limit
evolving as $L^*$ with redshift: $M_B<-18.5-2(U-B) - 1.3(z-0.9)$.}

\end{deluxetable*}

The intrinsic scatter of 0.18 dex in \logsigoned\ converts to 
$\sim 1.5$ magnitudes of intrinsic scatter.  This scatter is 
significantly larger than in low-redshift Tully-Fisher relations; 
most of these are restricted in morphology, 
and have scatter from 0.25-0.55 mag
(e.g. Tully \etal\ 1998; Willick 1999; Sakai \etal\ 2000), although
Kannapan, Fabricant \& Franx (2002) find an intrinsic scatter of 
0.5--0.6 mag in $B$ for bright spirals only but 0.8--0.9 mag in a 
sample which includes dwarfs and is not pruned on morphology.
There are several reasons for our high scatter, including: lack of
inclination and extinction corrections; our use of linewidth instead of 
circular velocity; and an all-inclusive sample that is broader than 
low-redshift samples, which are often highly selected to favor undisturbed,
orderly inclined disks.  However, high-$z$ Tully-Fisher morphologically
selected and inclination-corrected
rotation curve samples also show larger scatter than at low-$z$,
discussed further in Section \ref{sec-othertf}.


\subsection{Tully-Fisher relation for rotation curves}
\label{sec-rctf}

We can also construct Tully-Fisher relations using the
spatially resolved measures of velocity and dispersion,
\vrot\ and \sigtwod, measured with the {\sc ROTCURVE}
program on a subset of the 2-d spectra as discussed in Paper I.  
Paper I demonstrated that there are rotation and 
dispersion dominated galaxies (RDGs and DDGs), in which \vrot\ or
\sigtwod\ is respectively more important.  The dispersion
\sigtwod\ can represent disordered kinematics or other
velocity variations on scales below the seeing limit.
For the combined velocity measures
\wone\ and \whalf, where $S_K^2 = K\vrot^2 + \sigtwod^2$,
fitting yields TF results that are similar to the 1-d dispersion 
TF relation and its evolution,
albeit noisier since the sample is one-third as large.
The similar TF results are expected since \wone\ and \whalf\
correlate well with \sigoned.  The RDGs and DDGs show similar 
residual trends with redshift; at $z \sim 1$ both are 
shifted brighter than their low-redshift counterparts.
There is a small residual in the \sigoned\ Tully-Fisher relation 
as a function of $\vrot/\sigtwod$ or RDG/DDG-ness, discussed 
further in Section \ref{sec-evolresid}.

\begin{figure*}[ht]
\begin{center}
\includegraphics[width=5.5truein]{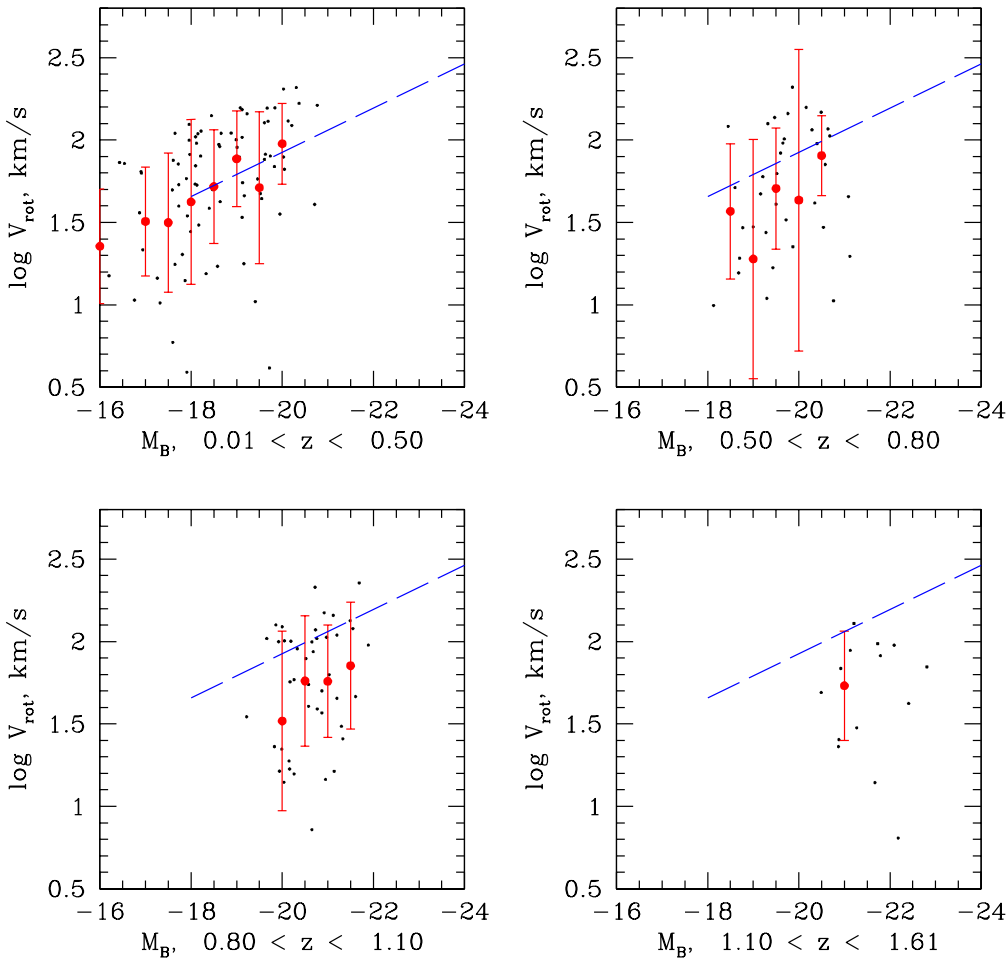}
\caption{The Tully-Fisher relation in the TKRS for 
line-of-sight rotation velocity and rest $B$ magnitude, in four
redshift ranges, for the galaxies in the ROTCURVE sample with 
aligned slits and ellipticity $e>0.25$.  Corrections for 
inclination and extinction are not applied.
Individual galaxies are plotted as small
points.  Large points and error bars are the mean
and RMS in magnitude bins.  The dashed diagonal line is
the fit to the low-redshift range, repeated in all four panels.
As in the linewidth-magnitude relation, at high redshift 
galaxies are observed brightward or lower velocity compared
to the low-redshift fit.
}
\label{fig-rctf}
\end{center}
\end{figure*}

The Tully-Fisher relation measured in the line-of-sight rotation 
velocity \vrot\ is also of interest.  Although \vrot\ does
not capture the full kinematic support of DDGs, the 
\vrot\ TF relation is most analogous to rotation velocity
measurements made in other works.  Figure 
\ref{fig-rctf} shows the $B$-band TF relation for \vrot, restricted to 
galaxies with ellipticity $e>0.25$ and aligned slits.  There are
not many galaxies in this sample, and we are only brave 
enough to fit a TF relation in the lowest redshift bin;
the low-$z$ fit is repeated in the other bins for comparison.
The low-$z$ fit has $A_{B,Vrot}=2.060 \pm 0.145$ and 
$B_{B,Vrot}=-0.134 \pm 0.081$, with covariant errors.  Its slope is 
consistent with local Tully-Fisher relations, and its intercept is
marginally lower in velocity, after compensating for the lack of
inclination and extinction corrections.  At least some of this 
offset is likely due to galaxies with low \vrot\ that are not dominated
by orderly rotation and would be excluded from local TF samples.  We have 
applied no inclination correction in Figure \ref{fig-rctf}, but 
since $e>0.25$, $\sini>0.66$, and the range of possible corrections
is limited.  The low-velocity objects in Figure \ref{fig-rctf} are 
something other than face-on disks.

As in the 1-d linewidth-magnitude TF relation, high-redshift points 
fall below or to the right of the low-$z$ \vrot-magnitude relation.  
The range of velocities observed is similar at low and high 
redshift, suggesting that the evolution is more in the
sense of brighter galaxies at high redshift, rather than a 
lack of high-velocity galaxies at high $z$.  The similar
behavior in the 1-d linewidth \sigoned\ relation provides evidence
that its evolution is real rather than a changing property
of the relation between 1-d linewidth and rotation velocity.

The scatter in this TF relation for rotation curves is quite
large, $\sim 0.3$ dex.  Inclination correction does not reduce 
it much because the sample was already restricted to $e>0.25$,
so the range of possible correction to velocity is 0 to 0.18 dex.
The scatter is high not because individual measurements
are bad, but because the sample has not been pre-selected 
to include only morphologically ordinary, orderly rotating disks.
A significant amount of the scatter is due to low-velocity
objects.  These are dispersion-dominated galaxies (see Paper I) 
with a component of kinematic support due to random or
disordered motions below the seeing limit.  Because they are
not primarily supported by an observed rotation gradient,  \vrot\ 
underestimates their kinematic support.  However, simply eliminating
DDGs from a TF sample is fraught with danger because it 
preferentially excludes low-velocity galaxies, introducing
a selection on \vrot.

\subsection{Testing fitting methods: fitting, scatter,
 and incompleteness biases}
\label{sec-fittest}

Methods for fitting linear relations such as Tully-Fisher 
have sometimes been controversial.  Several biases can arise, 
caused by observational errors, intrinsic scatter, and
magnitude limits.  Biases can be exacerbated in the
high-redshift TF relation by the large intrinsic scatter.
Here we describe tests of several fitting methods, to
insure that the parameters listed in Table \ref{table-bestpars}
are reliable.

\subsubsection{Scatter-induced bias in least-squares fitting algorithms}

A common method of fitting a linear relation to data with errors 
in both coordinates is the {\tt fitexy} least-squares routine, 
derived from a $\chi^2$ minimization (Press \etal\ 1992).  
The {\tt fitexy} method does not model relations
with intrinsic scatter, so yields formally rejectable fits 
with $\chi^2/N>>1$, and can yield biased results when there is
scatter.  A method that does account for scatter was proposed by
Akritas \& Bershady (1996), but this method has been criticized by
Tremaine \etal\ (2002) and Novak \etal\ (2005).  These authors in turn
generalized the {\tt fitexy} method by adding the intrinsic scatter as an
effective error term in one of the coordinates, so that the best fit
has $\chi^2/N=1$.  In the Appendix
we show that this intuitive treatment (generalized least squares, or
GLS method) is derivable from a maximum likelihood model, a special case 
of the maximuum likelihood (MLS) method we have used.  None of these 
models explicitly compensate for effects caused by selection limits.

To test biases introduced by scatter and selection limits,
we generated simulated data sets with
Monte Carlo realizations.  We took the true values of $M_B$ in a
single redshift range, 
enforced a TF relation with slope $B_{model}=-0.1$, and perturbed the
points by gaussian random variates of the observational errors
err($M_B$), err(\logsigoned), and an intrinsic scatter $C$ 
in \logsigoned\ of 0.15 dex.  Measuring and applying the 
intrinsic scatter in \logsigoned\ rather than in $M_B$ is required 
because the sample is magnitude selected, and sensible because 
the slope is shallow.

As the scatter $C$ is increased, bias in some of the methods
increases to be quite significant.  For the {\tt fitexy} method
without accounting for intrinsic scatter, the fitted slope
was typically $B_{model,fit} \sim -0.15$.  We believe that the algorithm
increases the slope to compensate for the additional scatter-induced
dynamic range of the sample in \logsigoned.  However, the
generalized least-squares GLS routine with scatter $C$ added in
quadrature to err(\logsigoned) had no measurable bias in slope.
The maximum likelihood MLS method described in the Appendix
is nearly identical and also tested free of bias.
The BCES$(Y|X)$ and BCES Orthogonal methods of Akritas \& Bershady (1996)
had a small bias, typically returning slopes $B_{fit} \sim -0.095$.

A sometimes-popular class of methods, the bisector fits (see e.g.\
Isobe \etal\ 1990;
used for Tully-Fisher in Ziegler \etal\ 2002; B{\" o}hm \etal\ 2004)
tested out to be extremely bad for this TF dataset when applied
without regard for magnitude selection.  Bisector
fits perform two standard least-squares fits using one set of errors 
at a time, fitting $(Y|X,err(X))$ and $(X|Y,err(Y))$ to obtain two fit lines,
and take the lines' bisector as the best fit.  Two problems with this
approach are: the two fits can have radically different $\chi^2$,
but bisecting weights them equally; selection limits can strongly
bias one of the fits, especially if the intrinsic scatter is large.  

In our dataset, because (1) there
is large intrinsic scatter; (2) the errors on $Y=\logsigoned$ are larger 
than on $X=M_B$; and (3) most of all because the sample is highly selected
in magnitude, the fit of $Y$ on $X$ is fairly good, while the fit
of $X$ on $Y$ is terrible in $\chi^2$ and in accuracy to the underlying
relation.  A typical $Y$ on $X$ fit yielded $B_{yx} =-0.098$, nearly
correct,
while the $X$ on $Y$ fit gave slope $1/B_{xy}=-1.18$, where $-10$ is
correct.  The bisector of these has slope $B_{bi,yx} = -0.42$, where
-0.1 was correct; the effect of the bisector was to mix a bad fit
in with a decent fit.  Essentially, in a magnitude selected
sample, determining the mean magnitude at a given velocity is only
meaningful if the scatter and the velocity errors are very small,
which is not the case in high-redshift TF samples.  When the
scatter is large, at a given velocity selection truncates the magnitude 
distribution, inducing a bias similar to Malmquist bias.

Accounting for intrinsic scatter within the fit method has two major
effects.  It removes biases such as the bias in slope we found
for the unmodified, no-scatter {\tt fitexy} method.  Adding scatter also 
reduces the weight given to individual measurements with small error bars.
For example, a measurement with log velocity error 0.06 dex that
deviates from our mean relation by 0.18 dex is a 3-sigma outlier
when intrinsic scatter is neglected, but only a 1-sigma deviation
given the intrinsic scatter.

An interesting consequence is that when a relation has large intrinsic
scatter, sampling variation has a stronger effect on the observations
than measurement noise.  Measuring a TF relation that has significant
scatter with a small number of non-noisy measurements is not efficient
and can be misleading.  Improvements in constraining a high-scatter TF
relation come from either a larger sample or decreasing the intrinsic
scatter, if possible.  In some sense this is the familiar problem of
overcoming cosmic variance by counting very large samples, but here
applied to a regression problem rather than density estimation.

\subsubsection{Incompleteness bias due to magnitude selection}

The simulations described above test biases in the fitting methods
due to scatter and due to the restriction in magnitude,
but they do not rule out a type of bias induced by sampling in the presence
of a magnitude limit and magnitude errors.
The ``incompleteness bias'' in truncated samples has been discussed in 
the local Tully-Fisher relation (e.g. Teerikorpi 1987; 
Willick 1994; Giovanelli \etal\ 1997; 
Tully \& Pierce 2000), but less so for high-redshift samples.
This bias is potentially important when using 
the forward TF relation as a distance indicator, since fitting 
magnitude on velocity is strongly affected by the magnitude limit, 
but that is not the case here since we use the inverse TF relation.  

If errors in magnitude are zero, the bias is eliminated by fitting 
velocity on magnitude (Schecter 1980; Willick 1994; Tully \& Pierce 2000).  
Our magnitude errors are non-zero, nearly ignorable but not quite.
The bias can also be induced by inclination corrections which make velocity
and magnitude covariant, but we do not apply those.  
One way this bias could affect the measurements is that faint galaxies that 
are scattered into the sample by observational errors tend to have lower 
velocities than slightly brighter galaxies that scatter out of the sample.
Another possible effect is if the fitting methods which take into
account magnitude errors react badly to the
truncation of the data.

In practice, the most straightforward way to calibrate incompleteness
bias is through Monte Carlo simulation (e.g. Giovanelli \etal\ 1997).
We constructed samples by extending the magnitude distribution below 
our magnitude limit, forcing a TF relation with slope $B=-0.1$,
applying the intrinsic scatter and observational errors, 
truncating the sample at the magnitude limit, and refitting.  In
the methods that treat scatter as an effective $Y$-error, incompleteness
bias was undetectable relative to the variation among Monte Carlo samples,
much smaller than the error estimates on $A$ and $B$.  The bias is
small because the magnitude errors of $\sim 0.12$ mag are small compared
to the 2-3 magnitude range of the sample, and the slope of velocity
on magnitude is shallow.

\section{Comparison to other Tully-Fisher measurements}
\label{sec-othertf}

Several works have previously measured Tully-Fisher
evolution, through either linewidths or rotation curve
modeling.  The picture has been confusing since methods,
samples, and the amount of TF evolution found are different.
Linewidth studies, each of $\sim 20$ galaxies, have generally found about
1-2 mag offset from the local Tully-Fisher relation at $z\sim 0.5$
(Forbes \etal\ 1996; Rix \etal\ 1997; Mallen-Ornelas \etal\ 1999).
These results are difficult to interpret because the samples
often were selected for high emission and because they were
calibrated by reducing a local TF relation to linewidths,
rather than internally.  Pisano \etal\ (2001) suggested that
these high-emission galaxies are vulnerable to linewidths
that underestimate $V_c$;  we have tested for this effect
with our spatially resolved fits in Paper I, and find little evidence
for it in a sample that spans the whole of the blue population.

Rotation curve studies of DEEP 1 galaxies (Vogt \etal\ 1996, 1997; 
Vogt 2000; Conselice \etal\ 2005) found fairly little evolution, $<0.5$ mag
in $B$ and $<0.3$ mag in $K$, to $z\sim 0.7$.  Flores \etal\ (2006) found 
little evolution in $K$ in a sample highly selected on rotation
properties.  In contrast, Simard \& Pritchet 
(1998) found $\sim 1.5$ mag evolution in $B$, in a sample selected 
to have moderately high emission EW, with median $z \sim 0.4$.
Bamford \etal\ (2005, 2006) found an evolution of $1.0 \pm 0.5$ mag
in $B$ projected to $z=1$ in a sample of 89 galaxies with median
$z=0.4$.

Studies in the FORS Deep Field (FDF) claim to find relatively little 
evolution at high mass and 1-2 mag at low mass (Ziegler \etal\ 2002; 
B{\" o}hm \etal 2004).  This is opposite the sense of differential evolution
that we find; their high-$z$ slope is steeper in $V(M)$, which is 
strongly rejected by our data.  However, Kannappan \& Barton (2004) 
argue that the evolution seen by the FDF group at the faint end is caused 
by kinematically anomalous galaxies.   Bamford \etal\ (2006)
show that the magnitude residuals are skewed in the FDF
magnitude-limited sample, meaning that only the brighter low-mass
galaxies are seen at high redshift. 
For this reason, the bisector method used in the FDF group's TF fits
yields biased slopes
for samples with large scatter, as argued in Section \ref{sec-fittest}.
We performed MLS fits to the 77-galaxy sample of B{\" o}hm \etal\ (2004)
and find inverse-TF slopes of 0.17 and $0.19 \pm 0.03$ dex/mag for
the $z<0.5$ and $z>0.5$ galaxies respectively, with $0.4 \pm 0.3$
mag evolution in intercept (from median redshift 0.7 to 0.3), 
and 0.12 dex intrinsic scatter in log $V_c$, in general agreement
with the fit of Bamford \etal\ (2006) to the full B{\" o}hm \etal\ (2004)
sample.  These slopes are closer to the local value than the bisector 
fits were, and there is no evidence for slope evolution within the sample.

The amount of evolution that we find, $\sim 1.0-1.5$ mag evolution 
in $B$-band from $z=0.4$ to 1.2, is larger than many of the rotation-curve
samples.  In part this may be because we have a larger number
of galaxies at high redshift; although most of the rotation curve
samples extend to $z=1$, their median redshifts are $\sim 0.5$.
Another more significant issue is that the linewidth 
sample has fewer selection effects and can incorporate 
morphologically and kinematically unusual objects, especially 
the dispersion-dominated galaxies.
Our sample is, in the mean, brighter than that of Vogt \etal\
(2000), and it is the brightest galaxies that show the 
most magnitude evolution.  We speculate that the amount
of luminosity evolution and degree of kinematic peculiarity
could be linked.


Our linewidth TF relation has an intrinsic scatter of 0.18 dex or 
about 1.5 mag in $B$-band, which is large but about equal to
the scatter predicted just from random inclinations of 
pure circular rotating disks (Section 4.3
of Paper I).  However, as discussed in Section \ref{sec-tfzbins},
the agreement with this inclination-induced scatter is fortuitous.
Applying inclination and extinction corrections does not reduce
the scatter.  As suggested in Paper I, there could be many galaxies
in the sample for which the kinematics are non-disky or the
ellipticities yield misleading inclinations.

In fact, the contribution of dispersion or 
disordered motions to integrated linewidth probably reduces the 
inclination-induced scatter, because galaxies that are not ideal
disks and have disorderly motions are less likely to fall to very 
low velocity when viewed face-on.  However, the disordered motions
may increase scatter induced by kinematic peculiarities, because 
the relation of the observed velocity to the halo mass is less
direct.   For inclination-corrected
rotation curve samples, the Kannappan \etal (2002) local TF,
which encompasses a wide range of galaxy types, has intrinsic scatter 
of 0.8--0.9 mag.  Our fits to the B{\" o}hm \etal\ (2004)
sample have an intrinsic scatter of 0.12 dex or $\sim 0.7$ mag.  The
rotation curve sample of Vogt \etal\ (2000) and Conselice \etal\ (2005) 
has a scatter of 0.7--1.1 mag even in $K$ band, which usually has lower 
scatter than $B$.  The subsample of Flores \etal\ (2006) that is
restricted to orderly rotating galaxies has remarkably low
scatter; these authors advocate the restriction of high-redshift
TF samples to only orderly galaxies, but this would devalue the
TF relation as an indicator of the evolution of the full blue
galaxy population.  Their larger sample including kinematically
anomalous galaxies has a much larger scatter in \vrot, possibly in 
part because \vrot\ does not include the kinematic support
from dispersion or disordered motions, as argued in Paper I.  

In general, it appears
that high-redshift TF samples have larger intrinsic scatter than the local
TF relation, even when rotation curves are used.  When linewidths
are used and are not inclination corrected, the intrinsic scatter appears
to increase, but not as much as one might expect from an ideal
rotating disk model, because not all galaxies fit that model.

\section{Discussion: Tully-Fisher evolution and its causes}
\label{sec-evolcause}

The results of Section \ref{sec-lwtf} show that there are Tully-Fisher 
relations between line-of-sight integrated linewidth \logsigoned\ and 
magnitudes $M_B$ and $M_J$ and that these relations evolve with redshift.
The relations remain linear, with large scatter, within our
ability to measure them.  The data show a very strong detection
of evolution in TF intercept in $B$-band, little evolution in intercept 
in $J$, and a moderately significant
(3 sigma) evolution in the slopes.  These measurements of evolution
are internal to our sample; our $z \sim 0.4$ relations are 
fairly consistent, within the errors, with local TF relations 
after $W_{20}$ is converted to \sigoned.

\subsection{Selection effects versus real changes in galaxy properties}
\label{sec-evolselec}

Several factors could cause evolution in the observed relations.
The evolution could be induced by selection effects on the sample
beyond simple magnitude selection, which we discussed in the
previous section.  The evolution could occur 
in the properties of integrated kinematics, e.g. in 
the $\sigoned/\vrot$ ratio due to changes in galaxy 
velocity fields or emission distributions.  
Or the relation could reflect true evolution in the properties
of galaxies, either in luminosity; or in measured velocity,
which is determined by dynamical mass, radius, and
concentration; or both.  

\subsubsection{Sample selection effects on TF evolution}

In Section 3
of Paper I we showed that the sample
of galaxies with linewidths is drawn evenly from the 
blue galaxy population in the TKRS.  The fraction
of blue galaxies without linewidths is small, $<20\%$;
since the missing galaxies are not systematically biased,
they are too few to induce a significant bias in the
measured relations.  The differential evolution 
in the $B$-band TF relation means that the evolution is
strongest for the brightest, easiest to detect galaxies;
typical selection effects operate the opposite way.

The TF fits to the sample described in Section \ref{sec-fitsamples},
with a rolling magnitude limit matched to
the color-dependent magnitude limit at $z=0.9$, are
matched so that we are fitting to the same luminosity and color
range of galaxies, relative to $L^*$, over a wide range of 
redshift.  These TF fits, tabulated in Table \ref{table-bestpars},
are very similar to the $M_B<-18$ sample.  Thus the TF
evolution measurement is not very sensitive to the
location of the limit in the color-magnitude plane.

Because we are measuring the TF relation in galaxies
with emission, effectively only in blue galaxies,
one could imagine a selection effect caused by galaxies
moving from the blue to red population as time
increases.  This would be a selection bias caused
by a genuine evolutionary effect.  
The effect required to produce slope evolution is that 
among the brightest blue galaxies, preferentially the 
low-velocity ones would have to move to the red sequence.
Then the (inverse) TF relation would steepen with time.
This effect seems very unlikely.  
If anything, one would
expect the opposite: the more massive, high-velocity galaxies 
should be older and stop star formation earlier.

It is very probable that mass is built up
on the red sequence in part by galaxies which age, redden,
and move from blue to red between $z \sim 1$ and now
(Bell \etal\ 2004; Faber \etal\ 2006) and there are signs
that the brightest blue $z\sim 1$ galaxies are on average located in
dense environments, suggesting they will become red by
the present day (Cooper \etal\ 2006).
However, the blue galaxy luminosity function is consistent with 
an evolution in luminosity and relatively little evolution in 
number density in the same period (Willmer \etal\ 2006).
If the percentage by number of blue galaxies which leave
the sample is small, the effect of their departure on TF 
fitting is also small.  TF relations with yet larger numbers of
galaxies are needed to constrain the behavior of rare
subpopulations.

\subsubsection{Evolution in galaxy properties and residual correlations}
\label{sec-evolresid}

The measured Tully-Fisher evolution could be an evolution
in fundamental physical properties such as mass and luminosity,
or an evolution in the relation of observables to properties,
e.g. dispersion to rotation velocity, or rotation velocity 
to mass.  It has been suggested that increased star formation
in the centers of galaxies or compact galaxies at high redshift 
could cause integrated linewidths to underestimate rotation velocity
(Pisano \etal\ 2001; Barton \etal\ 2001).  
Measurements of a variety of local galaxies
(e.g. Kobulnicky \& Gebhardt 2000) and the simulations presented
in Section 4.2
of Paper I suggest that linewidth is
actually a fairly robust measure, and that underestimating
\vrot\ is only likely to be a problem in very extreme
star-forming objects.

Our sample provides an empirical test.
Figure \ref{fig-lwtf} shows that the evolution we measure in $B$-band
is if anything largest for the brightest galaxies.   The brightest
galaxies are also comparatively red even at high redshift, as the 
color-magnitude relation in Figure 3
of Paper I shows, while extreme star-forming objects are quite
blue.  It is hard to explain the measured TF evolution with an 
effect caused by extreme star formation episodes.  

\begin{figure}[ht]
\begin{center}
\includegraphics[width=3.5truein]{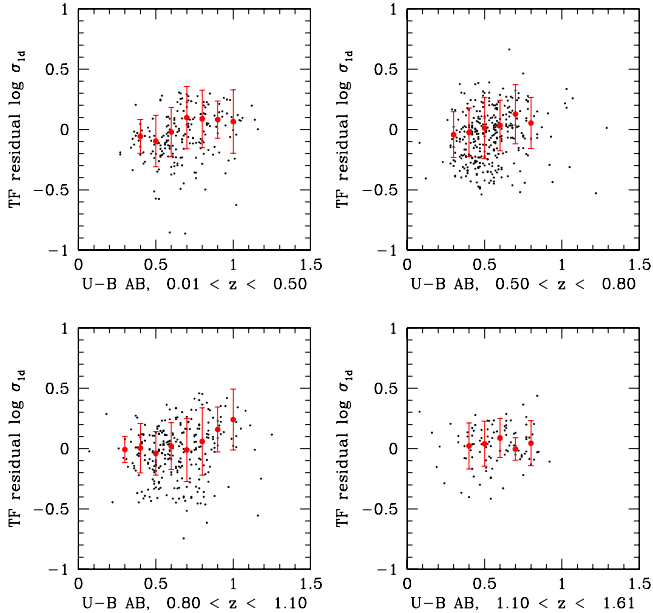}
\caption{Tully-Fisher velocity residual as a function of $U-B$ 
restframe color, for galaxies with $M_B<-18$.  
There is a weak correlation between
TF residual and color, with redder galaxies being at slightly higher
velocity, hence slightly subluminous.  
The sense of this correlation is such that measuring
the TF relation in a redder bandpass will decrease the residual.
The weakness and lack of evolution in the color-residual correlation
suggests that TF scatter and 
evolution are not driven by episodes such as extreme blue starbursts.}
\label{fig-colorresid}
\end{center}
\end{figure}

Another test comes from the color-TF residual relation in the $B$-band,
which is shown in Figure \ref{fig-colorresid}.  We compute the
TF residuals in velocity (\logsigoned), not magnitude, to avoid biases
caused by the magnitude limit.   There is a weak relation 
between $U-B$ color and TF residual, with redder galaxies
displaced to higher velocity, or equivalently lower $B$
luminosity.  The sense of the relation is such that
measuring the TF relation in a redder bandpass, more closely
related to stellar mass, will decrease the residual
(cf. Kannappan \etal\ 2002).  Highly obscured starbursts 
would have little effect on either $M_B$ or line emission,
though they might affect $M_J$.  Unreddened starbursts that cause
spuriously low linewidth of emission should be bluer and lie below the
mean relation in \sigoned.  Although there is a color-residual
trend, it is weak and evolves little with 
redshift, so it cannot be the primary driver of TF
evolution in magnitude, linewidth, or slope.  

\begin{figure}[ht]
\begin{center}
\includegraphics[width=3.5truein]{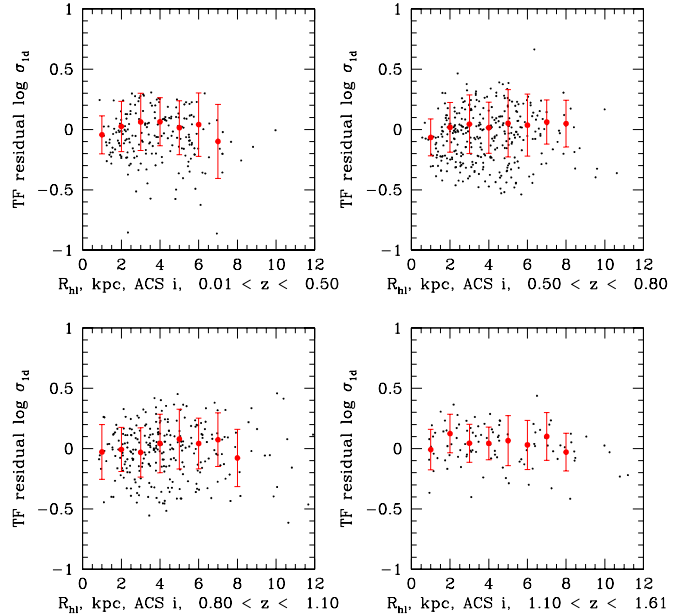}
\caption{Tully-Fisher velocity residual as a function of half-light radius
$R_{hl}$ in ACS $i$, for galaxies with $M_B<-18$.  
There is at most a weak correlation between
TF residual and radius, with larger galaxies being at slightly higher
velocity, hence slightly subluminous.  
The correlation, if any, changes little with redshift.  
}
\label{fig-radiusresid}
\end{center}
\end{figure}

Figure \ref{fig-radiusresid} plots $B$-band TF residual as a 
function of HST/ACS half-light radius.  This shows essentially
no correlation of TF residual with $R_{hl}$, suggesting that
TF evolution is not driven by compact galaxies alone.

\begin{figure}[ht]
\begin{center}
\includegraphics[width=3.5truein]{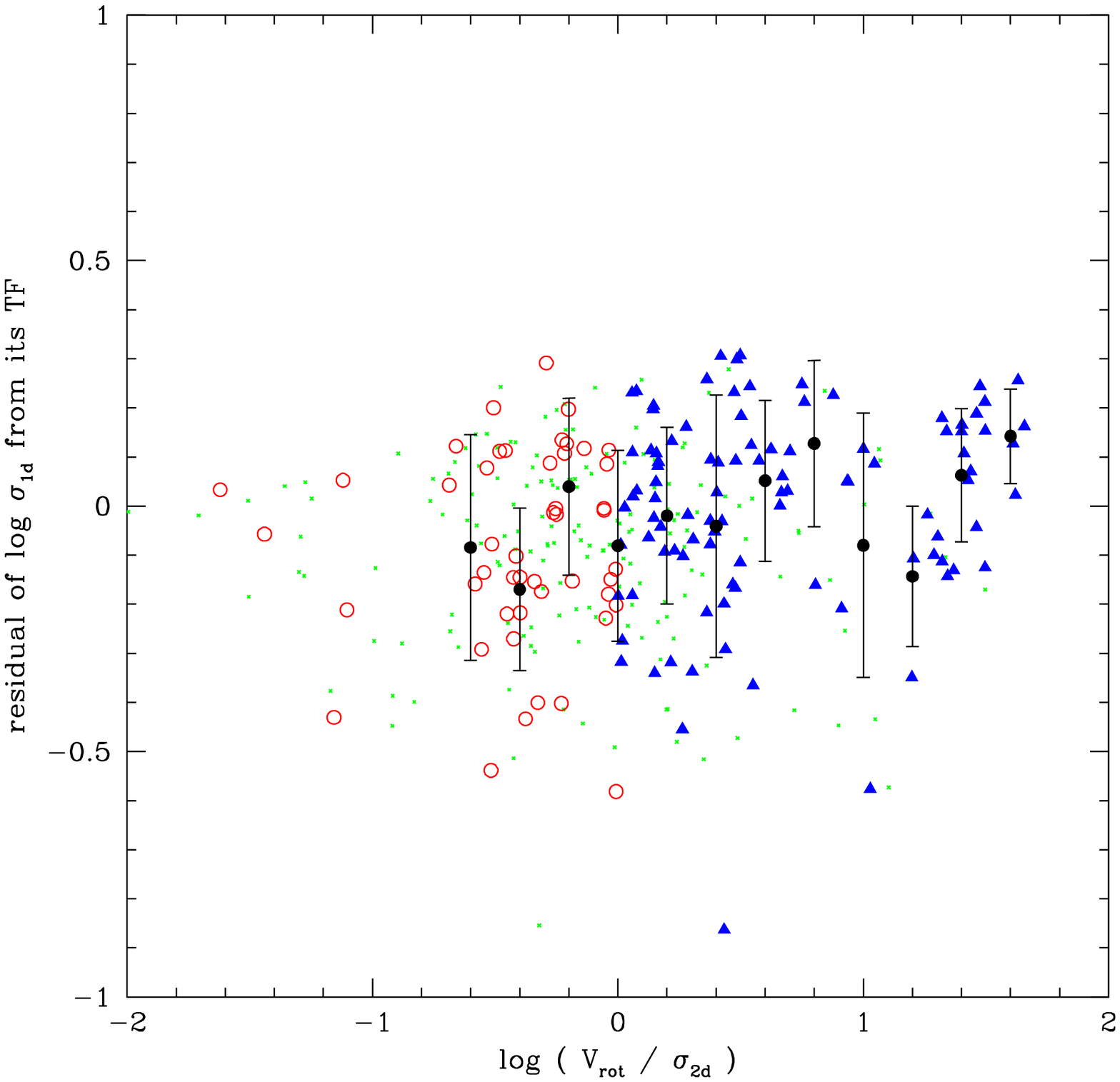}
\caption{Tully-Fisher velocity residual in \logsigoned\
as a function of velocity/dispersion
ratio measured in the 2-d spectra, ${\rm log}(\vrot/\sigtwod)$,
for galaxies with $M_B<-18$.  
As in Paper I, rotation dominated galaxies are plotted as filled
triangles, dispersion dominated galaxies are plotted as open circles,
and round galaxies or those with misaligned slits are plotted as small Xes.
The large points and errorbars are the mean and RMS of the 
aligned-slit galaxies.  There is a mild relation between TF residual
and $\vrot/\sigtwod$.
}
\label{fig-voversigresid}
\end{center}
\end{figure}

Figure \ref{fig-voversigresid} plots $B$-band TF residual in \logsigoned\
as a function
of the velocity to dispersion ratio $\vrot/\sigtwod$,
from the {\sc ROTCURVE} fits to the 2-d spectra.  Because the
number of galaxies with {\sc ROTCURVE} fits is smaller than the
whole sample, we combine galaxies at all redshifts on the same
plot, but we compute the TF residual from the fit appropriate
to the galaxy's redshift.  The triangles and open circles are
rotation and dispersion dominated galaxies with $M_B<-18$
measured in well-aligned
slits, as in the figures of Section 5
of Paper I.  The filled
circles and error bars are the mean of TF residual for the
well-aligned-slit galaxies.  There is a mild relation of 
residual with $\vrot/\sigtwod$, in the sense that dispersion
dominated galaxies are lower \sigoned\ for their magnitude.

We did not find evidence that this correlation with
residual changes with redshift.  Larger samples are needed to
study redshift evolution of the resolved kinematics in any detail
and to tell if evolution in the relative numbers of rotation and 
dispersion dominated galaxies could cause changes in the TF relation.
The trend in TF residual is much smaller than the RMS scatter,
indicating that the large scatter in the \logsigoned\ TF relation
is not caused simply by mixing rotation and dispersion dominated
galaxies.  It does appear that the most rotation-dominated 
galaxies, with high $\vrot/\sigtwod$, have smaller scatter about the
TF in \logsigoned.  These are likely to be the galaxies with the
most orderly kinematics.

\subsubsection{Evolution in velocity or luminosity?}
\label{sec-evolprops}

If the Tully-Fisher evolution is caused by an actual
evolution in velocity or luminosity, which is the more
likely suspect?  A galaxy's luminosity can evolve
substantially over an interval of several Gyr; we
discuss luminosity evolution models in Section \ref{sec-models}.
The fact that our
$B$ and $J$ Tully-Fisher relations evolve differently is a 
very strong argument that luminosity is evolving, since 
the aging of stellar populations causes a faster fading 
in $B$ than in $J$.

We know that galaxies can change substantially in luminosity
either through star formation episodes or fading, and that the blue
galaxy population fades by $\sim 1.3$ mag per unit redshift 
(Wolf \etal\ 2003; Willmer \etal 2006), which is similar to the 
evolution we find in Tully-Fisher intercept in terms of magnitude.
The locus of blue galaxies in color-magnitude space is consistent 
with this amount of quiescent evolution (Blanton 2005).  Measurements 
of the luminosity-metallicity relation suggest differential
luminosity evolution similar to what we found in the TF relation
(Kobulnicky \etal\ 2003; Kobulnicky \& Kewley 2004).  A study of the 
luminosity-size relation in the TKRS/GOODS sample suggests that there 
is differential luminosity evolution as a function of radius
with small galaxies fading more (Melbourne \etal\ 2006);
differential changes could be a function of both radius and mass.

A galaxy's characteristic velocity may also evolve with
time.  However, even in simplistic models, the relation between
halo mass and velocity evolves so that the increase in 
velocity as mass increases with time is small (Mo, Mao \& White 1998).
Essentially this is because the universe is denser at earlier times.
Simulations show that although halos are accreting mass as time goes 
on, the actual change in velocity dispersion is small at late times;
this is especially true for blue galaxies, which are less likely to have 
undergone a recent major merger.  A halo's central density is established 
early when its mass accretion rate is high (Wechsler \etal\ 2002).
As mass accretion slows, the characteristic central density begins
to asymptote, and later mass accretion builds up the outer parts,
so that the halo grows in outer radius, while the inner regions
where we measure velocity change little.  

A caveat is that because the inverse Tully-Fisher relation is
shallow, small changes in velocity could still be significant.
Nevertheless, our first step in modeling the TF evolution 
in Section \ref{sec-models} is
the minimal model: we consider galaxies' velocities as fixed
and impose pure luminosity evolution models.

\section{Discussion: Simple models for Tully-Fisher evolution}
\label{sec-models}

There does not yet exist a solid framework for predicting evolution
in the Tully-Fisher relation from fully fledged models of galaxy
evolution.  Numerical models of disk galaxy formation have 
reproduced some of the properties of the local Tully-Fisher
relation (e.g. Dalcanton, Spergel \& Summers 1997; Mo \etal\ 1998)
but using these to predict zeropoint and slope evolution has
many degrees of freedom since the run of $M/L$ with mass,
and its change with time, is relatively adjustable.

Using $N$-body plus gas-dynamical simulations, Steinmetz \& Navarro (1999)
predicted a TF intercept evolution of $\sim 0.7$ mag from $z=1$
to 0, but the evolution is highly dependent on the star formation
recipe, and the simulations do not match the local intercept
very well.  Recently, Portinari \& Sommer-Larsen (2006) have used
simulations to predict TF intercept evolution of $\sim 0.85$ mag 
in $B$ from $z=1$ to 0, but very little evolution in the stellar mass
TF intercept.  The $B$-band intercept evolution is moderately less than
we find, and the lack of stellar-mass intercept evolution agrees
reasonably with our $J$-band measurements.  However, they predict no
slope evolution, which does not agree with our data.
As in most $N$-body-derived models, their predicted low-redshift
TF intercept is offset from the local data, and the disk scale-lengths
are smaller than observed.  Portinari \& Sommer-Larsen correct
for the latter, which appears not to be a major effect, but these
differences indicate that it is still unclear how closely the evolution
in $N$-body models can be related to that of real galaxies.

Here we consider some simplified models for Tully-Fisher
evolution.  A major utility of the Tully-Fisher relation is that
it provides a way to link galaxies at different redshifts:
based on luminosity and color alone, it is difficult to say what a
blue galaxy at $z\sim 1$ will evolve into at low redshift
(but see Blanton 2005 for a treatment of the population
as a whole).  By using the additional dimension of characteristic 
velocity, we can relate, in the mean, a high-redshift galaxy to 
its likely descendants.  As a first simplification, as argued in 
Section \ref{sec-evolcause}, we suppose that the galaxies we
are measuring evolve solely in luminosity and that a given galaxy's
characteristic velocity as measured by \sigoned\ changes little over
our redshift range.  We assume that our measured TF intercept and slope
evolution are genuine, although the exact amounts are still uncertain,
and ask if they can be explained reasonably,
exploring the consequences of both intercept and slope
evolution for simple models of galaxy histories.

\subsection{Simple models: dwarfs and starbursting}

\subsubsection{Dwarfs: evolution at the faint end}
\label{sec-dwarfmodel}

One scenario that has been proposed is that the most active galaxies
at moderate redshifts are low-mass, faint, or dwarfs, and that
the strongest evolution should occur at the faint end of the
Tully-Fisher relation.  In this scenario, it is supposed that low-mass 
galaxies have had high star formation rates at redshifts $<1$
while high mass galaxies are already fairly evolved by then,
or at any rate that more evolution happens in faint galaxies at $z<1$
(e.g. Broadhurst, Ellis \& Shanks 1988).
Under this assumption, the luminosity evolution from $z \sim 1$ to 
now should be larger for low-mass galaxies.  This predicts that
at higher redshift, the inverse TF relation should be {\it steeper}.

Some intermediate-$z$ Tully-Fisher measurements have suggested this
sense of slope evolution, usually by detecting a small number
of bright galaxies
with low rotation velocities (e.g. Simard \& Pritchet 1998;
B{\" o}hm \etal\ 2004).  As discussed above, local samples suggest that 
kinematic anomalies could be responsible for some low velocity
measurements (Barton \etal\ 2001; Kannappan \& Barton 2004).
Further, the forward Tully-Fisher relation magnitude residuals at 
low velocity are skewed by the magnitude limit (see the Bamford \etal\ 2006
discussion of the B{\" o}hm sample),
since overluminous low-velocity galaxies can be observed but underluminous
ones are omitted from the samples.
This and the inappropriate slopes produced by forward and bisector fits,
discussed in Section \ref{sec-fittest}, are arguments for analyzing the 
inverse TF relation, i.e. velocity residuals as a function of magnitude,
as we do in this paper.

While the fast-evolving low-mass galaxy scenario sounds reasonable,
in fact the fits of Section \ref{sec-lwtf} show that 
the $z \sim 1$ inverse TF relations are marginally 
{\it shallower} than local.  A steepened relation in which, for example, 
low-mass galaxies are 2 magnitudes brighter while high-mass
galaxies are 1 mag brighter is definitively ruled out.

\subsubsection{Bursts: intermittent luminosity evolution}
\label{sec-burstmodel}

Another possible scenario is that galaxies are intermittently
brightened by bursts of star formation.  Locally, galaxies 
with high star formation rates per unit mass tend to be low-mass
(e.g. Brinchmann \etal\ 2004), but conceivably at higher redshifts, 
starbursting occurs in massive galaxies as well.  However, it is not 
clear that plausible starbursts are large enough to move the most
massive galaxies significantly (Barton \etal\ 2001).  In a simple
model of Tully-Fisher evolution driven by intermittent bursting,
galaxies move to brighter luminosity when a burst happens, but 
remain at roughly constant (or slightly lower) linewidth,
and then return to a baseline TF relation as the burst fades. 
To cause evolution, the bursts must be visible in $B$, rather than 
highly obscured starbursts.

This type of luminosity evolution driven by bursting has the desirable
effect of predicting a shallower slope for the inverse Tully-Fisher
relation at higher redshifts.  Effectively, the galaxy distribution
in $M_B - \logsigoned$ space is broadened in $M_B$, so that fitting
an inverse TF relation produces a shallower slope (although bursting
does not explain why the slope evolution is similar in $B$ and $J$).

However, if bursting is the major driver of TF evolution in
the $B$ band,  the most overluminous galaxies should be the bluest
in restframe color.  This is not true in our sample.  The 
color-magnitude relation for blue galaxies in Figure 3
of Paper I shows that even at $z=1$, the brightest blue galaxies 
are also the reddest of the blue galaxies.  The most luminous
blue galaxies have restframe $U-B_{AB} \sim 0.7 - 0.8$, typical of a 
blue disk with moderate star formation rate, such as a local Sb-Sc.
In Section \ref{sec-evolresid} we argued that the fact that the
$B$-band TF evolution is strong for bright, {\it moderate-color} galaxies 
excludes a starburst effect on linewidths as the cause; it also
excludes a starburst effect on magnitudes as the primary cause.

The weak relation of color and $B$-band Tully-Fisher residual shown
in Figure \ref{fig-colorresid} disfavors a bursting scenario.
If the Tully-Fisher
evolution were driven by more prevalent starbursts at higher redshift,
the color--TF residual correlation should be stronger at 
higher redshift.  Blue color and overluminosity (or low-velocity) 
should occur together, but Figure \ref{fig-colorresid} shows
that the color--TF residual is no stronger, and possibly weaker,
at higher redshift.
This is not to deny that starbursting happens or is more common
at high redshift.  Fluctuations in star formation rate could cause
some of the scatter in the $B$-band TF relation.  However, 
intermittent strong blue starbursts 
are not viable as the primary cause of Tully-Fisher slope 
evolution.

\subsection{Models parametrized by star formation history}
\label{sec-taumodels}

To make a more flexible and quantitative toy model, we calculate
luminosity evolution from simple star formation histories.
We assume that blue galaxies can be approximated by 
evolutionary tracks which build up stars over time,
either through mass accretion or gas consumption, and
use a family of models with exponentially declining
star formation rates, ``\taumods'' (Searle, Sargent
\& Bagnuolo 1973).  These are 
parametrized by the SF timescale $\tau$:

\begin{equation}
{\rm SFR}(t) \propto {\rm exp}(-(t-t_f)/\tau),
\end{equation}

\noindent
where $t_f = t(z_f)$ is the cosmic time at the formation
redshift $z_f$.  For a model which forms stellar mass $M_{*,tot}$ 
as $t \rightarrow \infty$,

\begin{equation}
{\rm SFR}(t) = \frac{M_{*,tot}}{\tau} {\rm exp}(-(t-t_f)/\tau),
\end{equation}

\begin{equation}
M_*(t) = M_{*,tot} (1 - {\rm exp}(-(t-t_f)/\tau)).
\end{equation}

Using \taumods, we make
star formation timescale a function of galaxy mass.
Models in which the star formation history depends on galaxy mass,
specifically so that massive galaxies undergo vigorous star
formation earlier than less-massive galaxies, 
have been much discussed recently under the rubric of
``downsizing'' (e.g. Cowie \etal\ 1996),
although the idea is of very long standing (e.g. Tully, Mould \&
Aaronson 1982).  Many studies have 
discussed the idea that SF timescale varies along the
Hubble sequence, which is also essentially a mass sequence
(e.g. Searle \etal\ 1973; Larson \& Tinsley 1978).

We used the stellar evolutionary code PEGASE (Fioc \&
Rocca-Volmerange 1997)
to compute luminosity and color as a function of time
for \taumods\ with a Kroupa IMF, solar metallicity, 
no stellar metallicity evolution, and with timescales 
$\tau = 0.1,2,4,8,10^4$ Gyr.  The 0.1 Gyr model is 
effectively a single burst model appropriate for red 
galaxies, and the $10^4$ Gyr model is effectively a 
constant star formation rate.  No extinction or correction 
for dust content is applied.  It is possible that 
evolving dust content affects the TF relation,
although the trends of increasing metallicity and 
decreasing gas fraction with time could offset each other.
Extinction must have less effect on the $J$-band TF relation than 
the $B$-band. 

\begin{figure}[htb]
\begin{center}
\includegraphics[width=3.5truein]{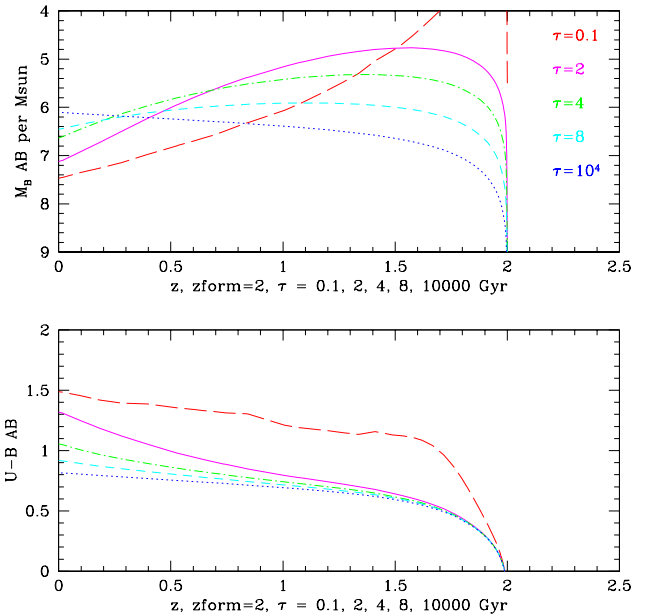}
\caption{Magnitude and color evolution in $M_B$ and $U-B$
of five $\tau$-models 
with $SFR \propto {\rm exp}(-t/\tau)$.  The models are started
at $z_f=2$, and have timescales $\tau=0.1,2,4,8,10^4$ Gyr.
The $\tau=0.1$ model is effectively a single burst and the
$10^4$ model a constant SFR.  The shorter-timescale models
fade more from $z\sim 1$ to the present; the constant-SFR model
gets brighter with time.}
\label{fig-ztautracks}
\end{center}
\end{figure}

\begin{figure}[htb]
\begin{center}
\includegraphics[width=3.5truein]{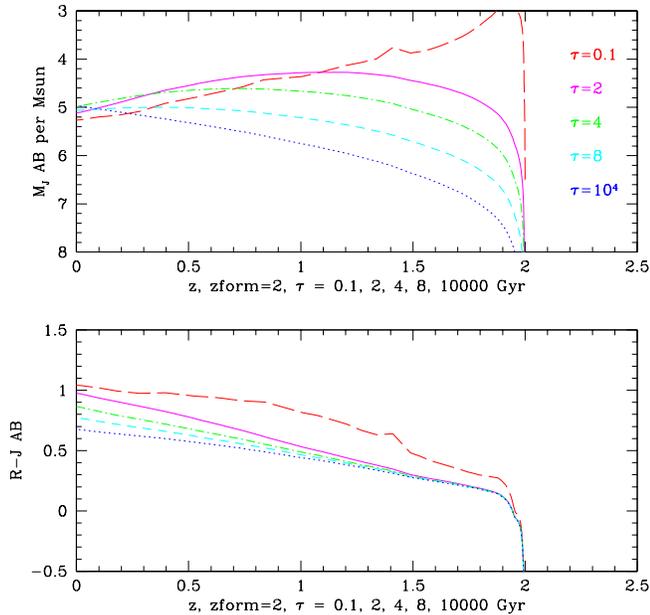}
\caption{Magnitude and color evolution in $M_J$ and $R-J$
of five $\tau$-models 
with $SFR \propto {\rm exp}(-t/\tau)$.  The models are started
at $z_f=2$, and have timescales $\tau=0.1,2,4,8,10^4$ Gyr.
The $\tau=0.1$ model is effectively a single burst and the
$10^4$ model a constant SFR.  The shorter-timescale models
fade more from $z\sim 1$ to the present, while the longer
timescale and constant-SFR models
get brighter with time due to the buildup of stellar mass.}
\label{fig-jztautracks}
\end{center}
\end{figure}

\begin{figure}[htb]
\begin{center}
\includegraphics[width=3.5truein]{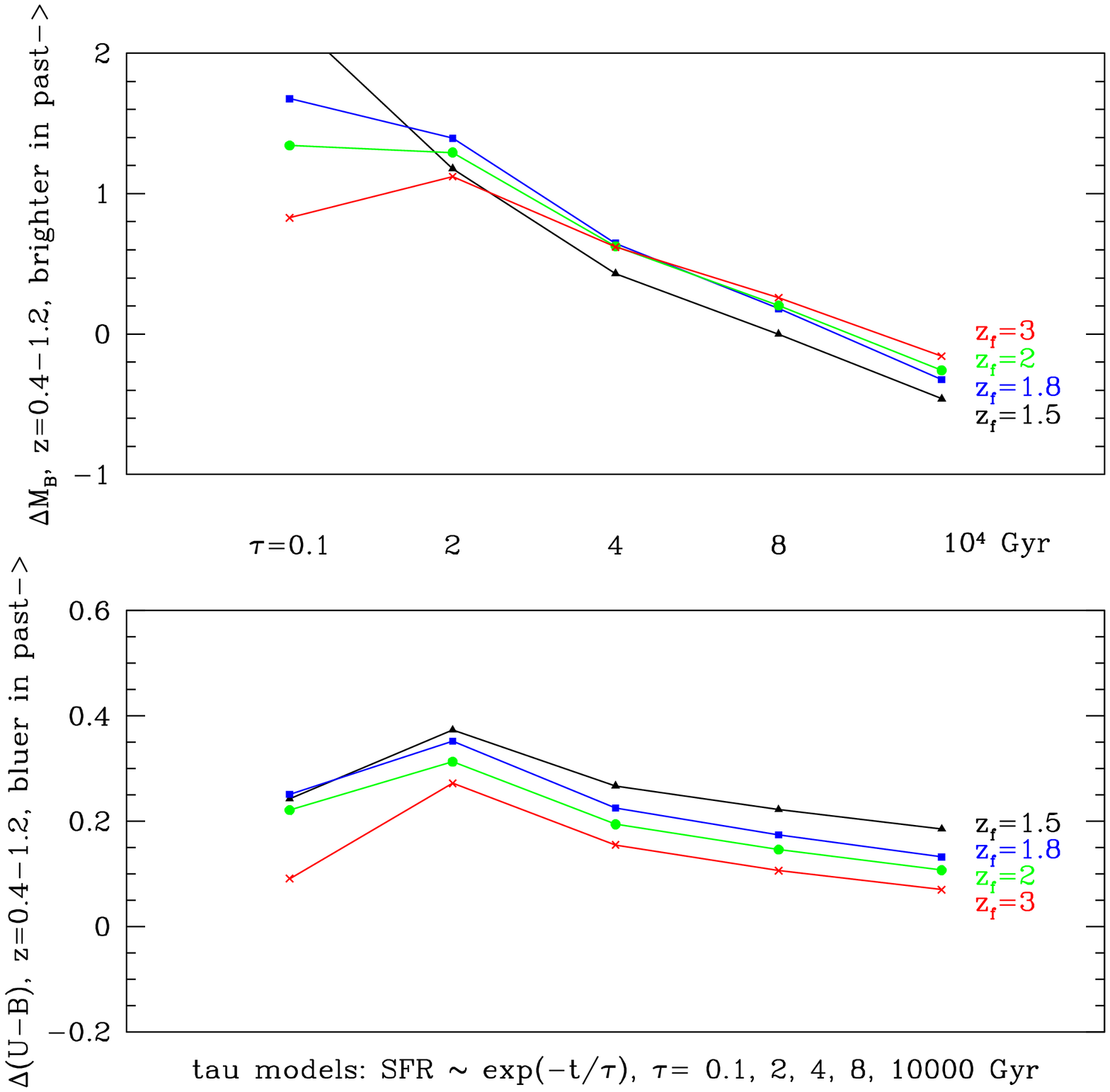}
\caption{Magnitude fading and and color reddening evolution in $M_B$
and $U-B$ of the
five $\tau$-models, with timescales $\tau=0.1,2,4,8,10^4$ Gyr.
The upper panel shows magnitude fading in $M_B$ from $z=1.2$ to 0.4 as
a function of timescale $\tau$, with different tracks for
formation redshifts from $z_f=1.5$ to 3.  The lower panel
shows reddening in $U-B$ in the same manner.
Fading and reddening are larger for shorter timescales (excepting the
$\tau=0.1$ single-burst model).  The amount of fading is strongly
dependent on timescale but only weakly dependent on formation redshift;
observations of fading in $B$ since $z=1.2$ do not distinguish 
between $1.8<z_f<3$.
}
\label{fig-taufade}
\end{center}
\end{figure}

\begin{figure}[htb]
\begin{center}
\includegraphics[width=3.5truein]{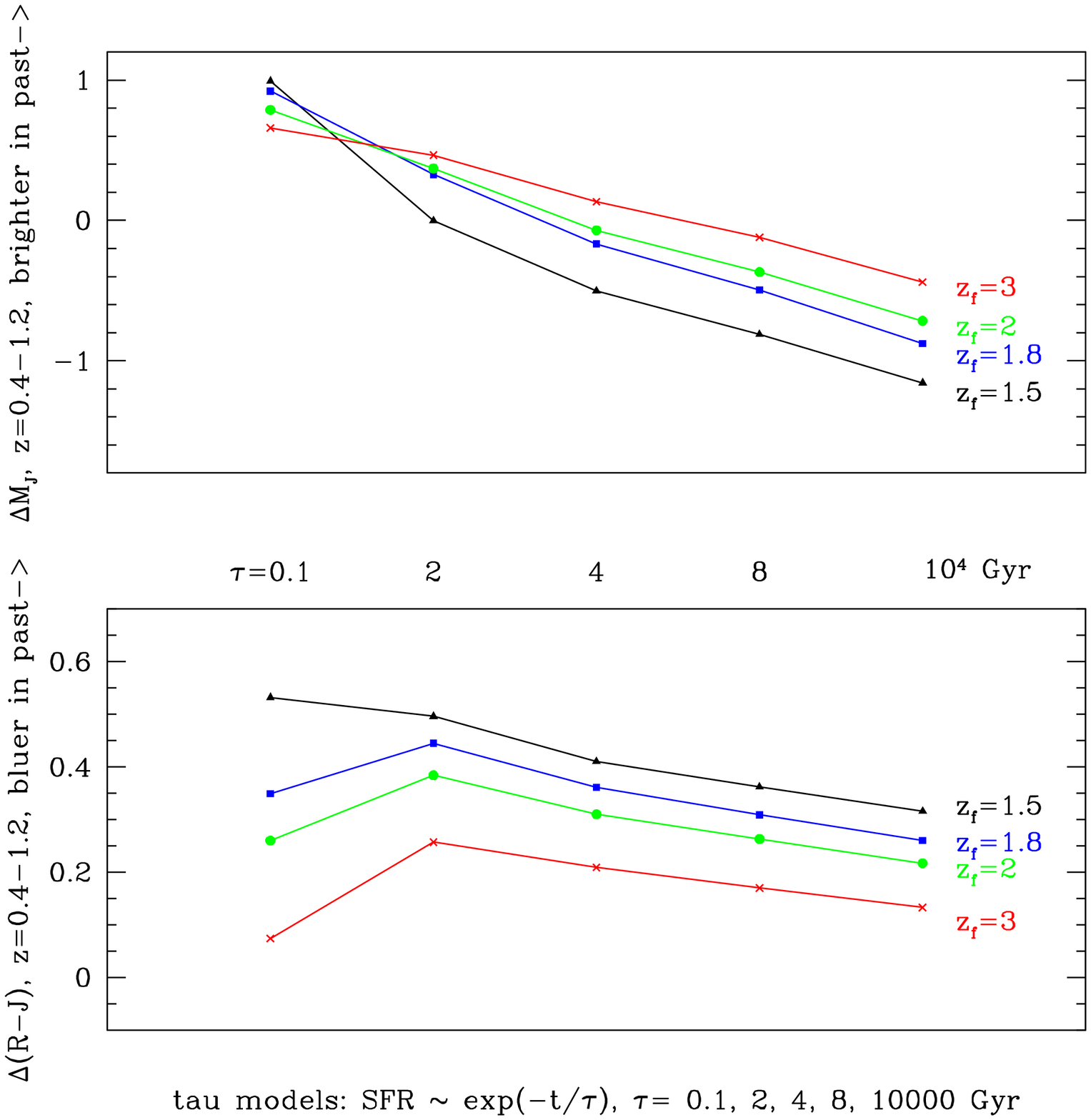}
\caption{Magnitude fading and and color reddening evolution in 
$M_J$ and $R-J$ of the
five $\tau$-models, with timescales $\tau=0.1,2,4,8,10^4$ Gyr.
The upper panel shows magnitude fading in $M_J$ from $z=1.2$ to 0.4 as
a function of timescale $\tau$, with different tracks for
formation redshifts from $z_f=1.5$ to 3.  The lower panel
shows reddening in $R-J$ in the same manner.
Fading and reddening are larger for shorter timescales (excepting the
$\tau=0.1$ Gyr single-burst model).  The amounts of fading 
and reddening are fairly strongly
dependent on timescale, and more sensitive to formation redshift
than are $M_B$ and $U-B$.
}
\label{fig-jtaufade}
\end{center}
\end{figure}

\subsubsection{Differential magnitude evolution in $\tau$-models}

Figures \ref{fig-ztautracks} and \ref{fig-jztautracks} 
show tracks of these models in
restframe magnitudes $M_B$ and $M_J$ (per unit total mass) and $U-B$ 
and $R-J$ colors,
for $z_f=2.0$.  The $\tau = 0.1,2,4,8$ Gyr models are normalized
so that they form 1 $M_\odot$ of stars as $t\rightarrow\infty$.
By comparing two different redshifts in the upper panel of
Figure \ref{fig-ztautracks}, we can see how much a given 
model fades, and thus how it evolves in the
Tully-Fisher relation.

The model track with $\tau=2.0$ Gyr fades by $\sim 1.2$ mag in $B$ from
$z=1.2$ to $z=0.4$.  Meanwhile, the model with $\tau=8.0$ Gyr
fades by only $\sim 0.2$ mag, and the constant star formation
track actually increases in brightness.  Similarly in $J$, the
short-$\tau$ model fades by about 0.4 mag, while the longer
timescale models brighten from $z=1.2$ to $z=0.4$.
In retrospect, this
behavior is easy to understand: short-$\tau$ models form a
substantial amount of their stars at high redshift, before we
have the chance to observe them, and these stars fade 
substantially from $z\sim 1$ to now.  In contrast, long-$\tau$
models are still building up a large fraction of their stellar
mass during the epochs we can observe, and fade little or even
increase in luminosity.

The near-single-burst model becomes very red in $U-B$ quickly and 
evolves only gradually after that.  The models with
$\tau \geq 2$ Gyr stay blue for a long time, gradually moving
redward, roughly consistent with color evolution in
the observed blue galaxy population (Weiner \etal\ 2005; Blanton 2005).
Toward lower redshift,
the $\tau=2$ Gyr model begins to peel away toward the red side
of the color bimodality.  However, because these are $U-B$
colors, even a small amount of additional late-time star
formation would bluen the color significantly; predictions
for $U-B$ color are less robust than predictions for $B$ 
magnitude.  The models only represent the galaxy population
in the mean, as the real blue galaxy population has a
color-magnitude relation with significant scatter in $U-B$ color
(e.g.\ Figure 3
of Paper I).

In Section \ref{sec-lwtf} we measured the linewidth Tully-Fisher
relation in four redshift ranges with median redshift from $z=0.4$ 
to 1.2.  Thus we ask how the models evolve from $z=1.2$ to 0.4.
Figures \ref{fig-taufade} and \ref{fig-jtaufade} show how the \taumods\
fade and redden, for different choices of $\tau$ and formation
redshift.  The amount of fading in restframe $B$ from $z=1.2$ 
to 0.4 is a strong function of $\tau$ -- apart from the 
single-burst $\tau=0.1$ model, the short-$\tau$ models
fade much more than the long-$\tau$ models.  In contrast, the
fading in $B$ is not a strong function of redshift of formation $z_f$,
unless $z_f$ is pushed very close to the epoch of observation.
However, the fading/brightening in $J$ is more sensitive to
$z_f$ than are the measurements in $B$.  In part this is because
change in $B$ measures relative change in the population of
young stars at a given epoch, while change in $J$ roughly
measures change in the integrated stellar mass formed from 
$z_f$ to the epoch of observation.
In reasonable ranges of timescale and $z_f$,
reddening in $U-B$ is not as strong a probe as fading.

\begin{figure}[htb]
\begin{center}
\includegraphics[width=3.5truein]{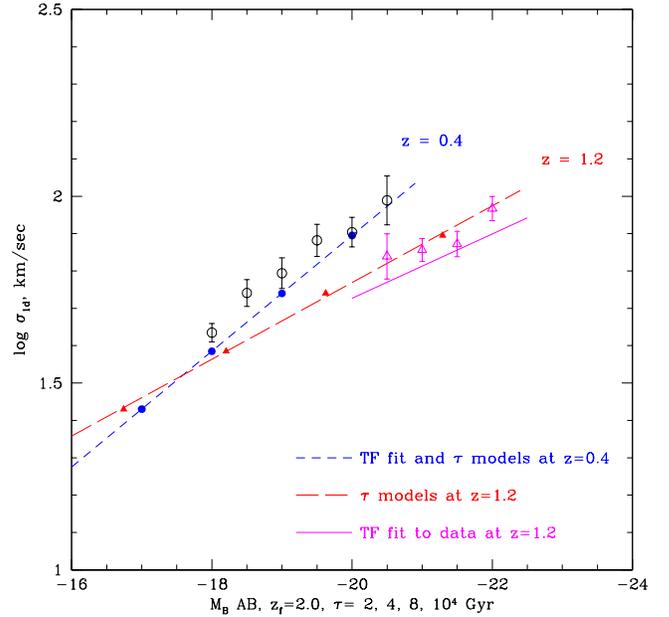}
\caption{$B$-band Tully-Fisher evolution predictions from $\tau$ models.
Four models with $z_f=2$ and $\tau=2,4,8,10^4$ Gyr (filled circles) 
are placed on the $z=0.4$ Tully-Fisher relation (short-dashed line).
The open circles are the $0<z<0.5$ linewidth data weighted means
binned by magnitude, and the error bars are standard error of the mean.
Data fainter than $M_B=-18$ are not used in the TF fit and are omitted.
The $\tau$ models are evolved 
back to $z=1.2$, assuming that \logsigoned\ does not change and
only $M_B$ evolves.  The $z=1.2$ models and a linear fit to them
are plotted as filled triangles and long-dashed line, showing the 
luminosity and slope evolution.  The $z=1.2$ models continue to
lie on a linear TF relation, although this was not forced by any
constraint.  The $1.1<z<1.61$ linewidth data 
binned by magnitude are plotted as open triangles, and the TF fit 
at $1.1<z<1.61$ is plotted as a solid line.}
\label{fig-tautfevol}
\end{center}
\end{figure}

\begin{figure}[htb]
\begin{center}
\includegraphics[width=3.5truein]{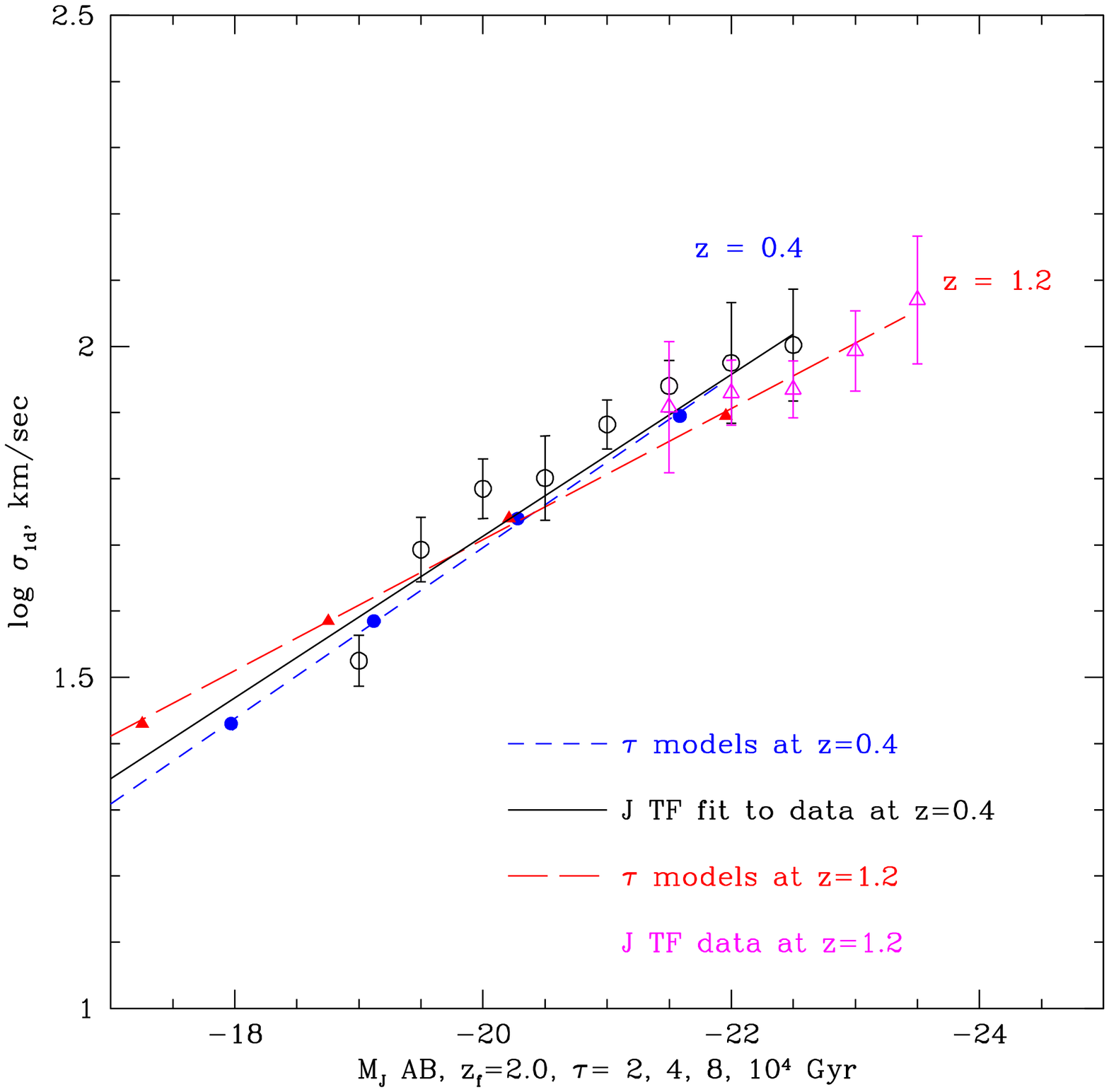}
\caption{$J$-band Tully-Fisher evolution predictions from $\tau$ models.
Four models with $z_f=2$ and $\tau=2,4,8,10^4$ Gyr
were placed on the $z=0.4$ $B$-band Tully-Fisher relation.  Their 
position in $J$ is then fixed by the models  (filled circles).
The open circles are the $0<z<0.5$ linewidth data weighted means
binned by $J$ magnitude, and the error bars are standard error of the mean.
Data fainter than $M_J=-19$ are not used in the TF fit and are omitted.
The $\tau$ models are evolved 
back to $z=1.2$, assuming that \logsigoned\ does not change and
only $M_J$ evolves.  The $z=1.2$ models and a linear fit to them
are plotted as filled triangles and long-dashed line, showing the 
luminosity and slope evolution.  The $z=1.2$ models continue to
lie on a linear TF relation, although this was not forced by any
constraint.  The $1.1<z<1.5$ linewidth data 
binned by $J$ magnitude are plotted as open triangles.}
\label{fig-jtautfevol}
\end{center}
\end{figure}

To show how \taumods\ can predict Tully-Fisher evolution,
we pinned each of the $\tau=\{2,4,8,10^4\}$ Gyr tracks to the
$z=0.4$ $B$-band Tully-Fisher relation.  We fixed $z_f=2$ and 
assigned these values
of $\tau$ to the magnitudes $M_B=\{-20,-19,-18,-17\}$ respectively,
so that shorter timescales belong to more luminous, hence more 
massive galaxies.
Once $z_f$, $\tau$, and magnitude at $z=0.4$ are chosen,
these parameters fix the normalization for each \taumod\ - 
effectively, the mass in stars+gas.  The \logsigoned\ which
corresponds to each $\tau$ comes from the $z=0.4$ TF relation.
Since we assume that \logsigoned\ does not change,
the model tracks then give $M_B$ at $z=1.2$ and the high-$z$
TF relation.

\begin{deluxetable*}{lrrrrrrrrrr}

\tablecaption{
$\tau$ models normalized to $z=0.4$ $B$-band TF relation
\label{table-taumods}
}

\tablecolumns{11}
\tablewidth{0pt}
\tabletypesize{\small}

\tablehead{
$\tau$, Gyr  &  $M_B$ &  $M_B$ & $M_J$ &  $M_J$ & \logsigoned & log $M_{b,tot}$\tablenotemark{a} & 
 log $M_*$\tablenotemark{b} & log $M_*$\tablenotemark{b} & log SFR & log SFR \\
        & $z=0.4$ & $z=1.2$ & $z=0.4$ & $z=1.2$ &   &   &  $z=0.4$ & $z=1.2$ & $z=0.4$ & $z=1.2$  \\
  &  &  &  &  & (\kms) & ($M_\odot$) & ($M_\odot$) & ($M_\odot$) & ($M_\odot$/yr) & ($M_\odot$/yr) 
 }
 \startdata
2       &  --20  &  --21.3  & --21.6 & --22.0 & 1.90  &  10.49  &  10.47  &  10.27  & --0.11 &   0.80 \\
4       &  --19  &  --19.6  & --20.3 & --20.2 & 1.74  &   9.99  &   9.88  &   9.55  & --0.27 &   0.19 \\
8       &  --18  &  --18.2  & --19.1 & --18.8 & 1.59  &   9.65  &   9.37  &   8.95  & --0.58 & --0.35 \\
$10^4$  &  --17  &  --16.7  & --18.0 & --17.3 & 1.43  &   9.28  &   8.76  &   8.24  & --0.92 & --0.92 \\

 \enddata

\tablenotetext{a}{Total stellar+gas mass of model}
\tablenotetext{b}{Stellar mass formed by the given redshift}

\end{deluxetable*}

\begin{deluxetable}{lrrrr}

\tablecaption{
Evolution in TF relation predicted by a set of $\tau$ models
\label{table-tfpredict}
}

\tablecolumns{5}
\tablewidth{0pt}

\tablehead{
Band & Redshift  & Zeropoint & Intercept A & Slope B   \\
     &           &     (mag) &         dex &   dex/mag 
 }
 \startdata
$B$  & $z=0.4$   & --21 & 2.050\tablenotemark{a} & --0.155\tablenotemark{a}  \\
$B$  & $z=1.2$   & --21 & 1.872 & --0.103  \\
\tableline
$J$  & $z=0.4$   & --22 & 1.954 & --0.129  \\
$J$  & $z=1.2$   & --22 & 1.906 & --0.099  \\

 \enddata

\tablenotetext{a}{Fixed to the observed $z=0.4$ $B$-band 
TF relation by construction.}

\end{deluxetable}

Table \ref{table-taumods} gives the magnitudes and masses
of the \taumods\ with $z_f=2$ normalized to the $z=0.4$ Tully-Fisher
relation.  Once each $\tau$ is assigned to a value of $M_B$ at $z=0.4$,
\logsigoned\ is fixed by the Tully-Fisher fit, $M_B$ at $z=1.2$ is
fixed by the past history of the \taumods, $M_J$ at both epochs
is fixed by the tracks of the \taumods, and the stellar masses
are given by their $M_*/L$.  For the $\tau=2,4,8$ Gyr models,
the total stellar+gas mass is given by integrating the SFR as 
$t \rightarrow \infty$.  The $\tau=10^4$ Gyr model does not converge
to a reasonable amount, so we set it to have a plausible
$M_*/M_{b,tot}$ at $z=0$.

Table \ref{table-tfpredict} gives the results of fitting ``Tully-Fisher
relations'' to the four \taumods\ once they have been assigned
values of $M_B$ and \logsigoned\ at $z=0.4$, by putting them
on the observed TF relation at $M_B=\{-20,-19,-18,-17\}$.  
The $B$-band TF relation at $z=0.4$ is the same as the
observed relation by construction.  The $B$ TF relation at
$z=1.2$ and both $J$ TF relations are determined by the 
tracks of the \taumods.  The TF relation of the models is
shallower at $z=1.2$ in both $B$ and $J$; the intercept 
evolution is much less in $J$ than in $B$.

Figure \ref{fig-tautfevol} shows the model TF relations compared
to the TF data in $B$.  The short-dashed line shows our $z=0.4$ TF relation
from Figure \ref{fig-lwtf}, and the \taumods\ which are pinned
to it are plotted as filled circles.  Rather than plot all
the galaxies, we show the weighted means of the $0<z<0.5$ data in
magnitude bins, as open circles; the error bars are
the standard error of the mean.  (The weighted means are higher
than the fit line because higher \logsigoned\ points have smaller
observational errors, as discussed in Section \ref{sec-dataprops}.)
The \taumods\ at $z=1.2$
are plotted as filled triangles, and the long-dashed line is a
fit to their \logsigoned\ on $M_B$.  The solid line is the TF relation 
fit to the galaxies at $1.1<z<1.61$, and the open triangles are the 
$1.1<z<1.61$ data in bins of magnitude.  The models produce
a high-$z$ $B$-band TF relation that is shifted and shallower
than at low redshift, as is seen in the data.

Figure \ref{fig-jtautfevol} shows the analogous comparison
between models and TF data in $J$.  Once the set of models is
normalized to the $z=0.4$ TF relation in $B$-band, all the
degrees of freedom are used up: the $z=0.4$ relation between
linewidth and $J$ magnitude is fixed by the model colors.
The $z=0.4$ model TF relation comes out linear and agrees quite well 
with the actual relation, which is encouraging, though it mostly
means that the models have the correct relation of $B-J$ color
to mass at $z=0.4$.  The models at $z=1.2$ produce a $J$-band TF
relation that is only slightly shifted in intercept, and
is shallower than at low redshift.  Essentially the
``pivot point,'' where the low and high redshift TF relations 
cross, is at higher mass (shorter $\tau$) in $J$ than in $B$-band.

The \taumods\ produce a magnitude evolution only slightly 
less than indicated by the high-redshift data and TF fit.  
Of course, we had a degree of freedom in picking
the normalization of $\tau$ to $B$ magnitude at $z=0.4$, but the
SFR timescales, colors and stellar mass fractions implied
are reasonable.  Thus the amount of
Tully-Fisher intercept evolution is consistent
with pure luminosity evolution and very reasonable star formation 
histories and colors.
The \taumods\ also preserve the linearity of the TF relations:
at $z=1.2$ they lie on a near-perfect line, though in principle
they could have produced a nonlinear relation.  This required no
fine-tuning, and again shows that the persistence of a linear
TF relation is consistent with a reasonable variation of SF 
history with mass.

The critical feature of using \taumods, with $\tau$ a function 
of magnitude or velocity, is that they naturally
produce differential fading, and pinning them along the low-redshift
TF relation produces a slope that will evolve with redshift.
Figure \ref{fig-tautfevol} shows that our normalization of \taumods\
to magnitude at $z=0.4$, which gives close to the correct
luminosity evolution, also produces slope evolution that agrees
with the data.

These \taumods\ are not at all the only plausible set of star formation
histories, and do not include metallicity or dust evolution;
observationally, there is still uncertainty in
the high-redshift Tully-Fisher slope, due to both small numbers
and the limited magnitude range at high redshift.  We do not claim that
the models are in any way unique.  However, Figures \ref{fig-tautfevol}
and \ref{fig-jtautfevol}
show that reasonable models for star formation history as
a function of mass naturally produce evolution in the 
Tully-Fisher ridgeline intercept and slope that matches both optical
and near-IR observations.

The \taumods\ also predict a color-magnitude relation and 
its evolution, although here metallicity evolution and dust 
require more careful treatment (Tully \etal\ 1982).
The differential evolution in luminosity and dependence
of $\tau$ on luminosity are generally consistent with
models derived from evolution in the luminosity-metallicity
relation (Kobulnicky \etal\ 2003; Kobulnicky \& Kewley 2004).
Ultimately, realistic models should be constrained by
evolution in the velocity-magnitude-color relations.
The models must also reproduce the scatter in these relations.
Our simplistic \taumods\ predict color and TF relations with
no scatter; in addition to metallicity and dust, it is almost
certain that episodic variation about the
smooth star-formation history produces some of the scatter
(e.g. Larson \& Tinsley 1978).

\subsubsection{Differential star formation rate evolution}


Because the high mass galaxies have the shortest SFR timescales
while low mass galaxies build up stars at a more nearly constant rate,
our simple model predicts that the $\sim 10\times$ higher
global star formation rate at $z \sim 1$ compared to local 
(e.g.\ Lilly \etal\ 1996) is dominated by higher SFR in 
{\it massive} galaxies, rather than low-mass objects.  
Our \taumods\ from high to low mass
have SFRs which are $\{1.6,0.8,0.4,0\}$ 
dex greater at $z=1$ than $z=0$, for $\tau=\{2,4,8,10^4\}$ Gyr.  
These are roughly compatible with the observed drop in SFR, although a 
full prediction of the evolution of the global SF rate requires a
convolution with the luminosity or mass function.  This scenario
agrees with $z=0.5-1$ observations of IR-luminous galaxies that 
show a high SFR in luminous spirals (Bell \etal\ 2005; Melbourne,
Koo \& Le Floc'h 2005).

\begin{figure}[htb]
\begin{center}
\includegraphics[width=3.5truein]{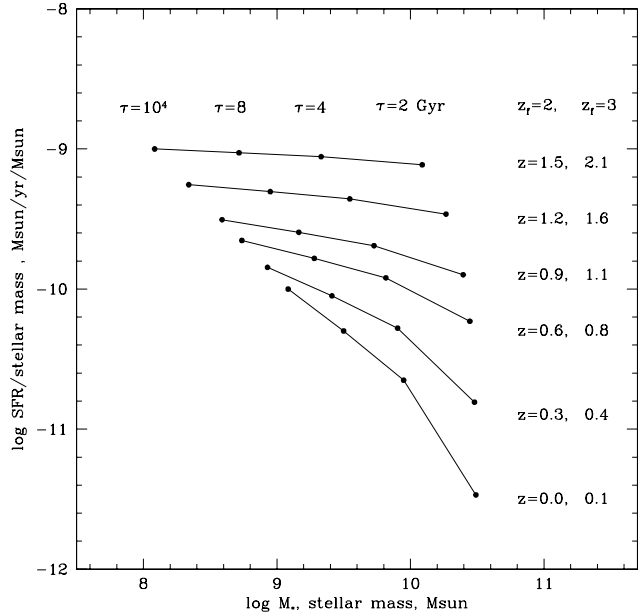}
\caption{An example prediction from \taumods\ of the 
evolution of the relation of specific star formation rate,
SFR/stellar mass, versus stellar mass.
The \taumods\ were started at $z_f=2$ and normalized
to fit the $z=0.4$ $B$-band TF relation.  Each set of
points connected by a line represents the specific 
SFR--stellar mass relation as a function of $\tau$
at a given redshift, labeled in the column of redshifts 
under ``$z_f=2$.''  Because high mass galaxies have shorter 
SFR timescales, they fall more quickly in specific SFR.  The 
rightmost column of redshifts, under ``$z_f=3$,'' shows where
the models fall for the same time gaps but a formation 
redshift of 3.  If the models are started earlier, the low-redshift
behavior is similar but the epoch where high-mass galaxies
have high specific SFR is pushed back.
}
\label{fig-sfrspecific}
\end{center}
\end{figure}

One interesting prediction of a model that makes SFR timescale 
a function of mass is that the slope of SFR on mass
should vary with redshift.  The low-mass galaxies with constant 
SFR evolve slowly, but the high-mass galaxies are at much higher SFR
at high redshift than now.
Figure \ref{fig-sfrspecific} plots the
evolution of our set of \taumods\ in specific SFR (SFR/stellar
mass) versus stellar mass.  The models are started at $z_f=2$ and
normalized to fit the $z=0.4$ $B$-band TF relation as
in the previous section.  Each set of connected points represents
the mean specific SFR-stellar mass relation at redshifts from 1.5
to 0.  The models predict substantial change in slope of this
relation.  However, the details of this relation
depend on the redshifts of formation.  The labels in the rightmost
column show the effect of starting the models at $z_f=3$; the
models reach a given evolutionary state at higher $z$.
Increasing $z_f$ can push back the epoch at which massive galaxies 
have high SFR, making them hard to observe,
and any real relation will have a large scatter.  If one pushes to
high enough redshift, massive galaxies may have multiple progenitors,
violating our assumption that a single track corresponds to a 
single object.  However, it is hard to avoid the general trend that 
high-mass objects or their progenitors form relatively
more stars at early times, in order to become older and redder by the
present day.

Measurements of SFR as a function of 
stellar mass ought to show a change in slope. 
Unfortunately, the limited depth of mid-infrared surveys
makes slope change difficult to see in 24 $\mu$m selected samples
(e.g.\ Bell \etal\ 2005).  There is considerable scatter in any
observed SFR--stellar mass relation, and selection effects tend to 
eliminate the low-mass and low-SFR objects at high redshift, making 
it hard to measure an unbiased mean relation.
Surveys of SFR from the \oii\ line
have tended to show a trend of SFR(mass) that does not 
evolve in slope, but rather an overall decrease in intensity with
time (e.g. Brinchmann \etal\ 1998; Bauer \etal 2005).  However,
SFRs from \oii\ uncorrected for extinction may have selection
effects and biases
that are correlated with mass and change with redshift.
Nevertheless, at some point, the SFR(mass)
slope must change with time, or it is very difficult to explain
the blue galaxy color-magnitude relation.

\section{Conclusions}
\label{sec-conclude}

We measured line-of-sight integrated kinematic linewidths \sigoned\ and 
spatially resolved line-of-sight rotation \vrot\ and disperson \sigtwod\
from the galaxy spectra of the Team Keck survey of the GOODS-N field.  
In Paper I we showed that linewidths are a measure of internal 
kinematics that is relatively robust against observational effects,
although there is significant scatter between any individual 
galaxy linewidth and true circular velocity or dynamical mass.  
We use 968 linewidths of galaxies with $M_B<-18$ 
and 677 linewidths of galaxies with $M_J<-19$ to measure 
evolution of the restframe $B$ and $J$-band Tully-Fisher relations
from $z\sim 0.4$ to 1.2.  
This is the largest sample of high-redshift galaxies
with kinematics to date, and samples the blue galaxy population
without morphological or other pre-selection.  
It allows both an internal comparison, without having to fix to a 
local TF relation, and the measurement of intercept and slope evolution.  

The intrinsic scatter in the Tully-Fisher
relation for linewidths is large, 0.18 dex in \logsigoned, which 
requires special care in fitting methods; we derive a maximum 
likelihood fitting method in the Appendix.  The intrinsic scatter
is partly due to the lack of inclination and extinction corrections;
kinematic pecularity is likely to both contribute to the scatter and
to reduce its non-gaussianity by reducing the number of orderly 
low-inclination objects with low observed velocity.  Correlations
of TF residual with other parameters are weak and do not suggest
obvious second-parameter sources of scatter or evolution.

In the $B$-band linewidth TF relation, there is very significant 
intercept evolution and $3\sigma$ evidence for slope evolution.  
The sense of the slope evolution is such that the high-redshift
inverse TF relation, velocity as a function of magnitude,
$V(M)$, is shallower than at low redshift.
The $J$-band linewidth TF relation shows relatively little evolution
in intercept, but a similar evolution in slope to that found in $B$-band.
The $B$-band TF relation for rotation velocities \vrot\ also supports 
intercept evolution, but the sample is too small and noisy to measure 
slope evolution.

The sense of slope evolution is most easily interpreted as differential
luminosity evolution: the most luminous galaxies at $z>1$ fade by
more than the less luminous galaxies.  There is more fading in $B$ 
than in $J$ band.  In the mean, a galaxy with $M_B=-21.5$ at $z=1.2$ 
fades by $\sim 1.5$ $B$ magnitudes to $z=0.4$, but only 
by $\sim 0.5$ mag in $J$.
The $B$-band evolution is larger than some previous TF measurements;
some differences are that our sample has many more galaxies
at high redshift, that it is not morphologically selected
to favor normal disks, and that it does not reject galaxies with
non-rotating kinematics.  Because the evolution is differential,
the amount of fading seen is also dependent on the magnitudes
of the different TF samples.  The fading of $1-1.5$ mag in $B$ at 
fixed velocity that we see is similar to the fading of $1.3 \pm 0.3$
mag in $L_B^*$ in the blue galaxy LF measured by Willmer \etal\ (2006),
and the fading of luminosity at fixed radius seen by Melbourne \etal\
(2006).  A simple, but not unique, model for these results is that
most of the blue galaxy population evolves unaltered by merging
or density evolution since $z \sim 1$, with stellar population
fading the primary driver of luminosity evolution.

We present a simplistic model to explain differential luminosity
evolution: we construct galaxy histories with star formation rate
declining exponentially with time 
(\taumods, ${\rm SFR} \propto e^{-t/\tau}$), and make
the SFR timescale $\tau$ a function of mass along the Tully-Fisher
relation.  In this scenario, massive galaxies have short $\tau$
and form the bulk of their stars early, before $z=1$, while
low mass galaxies have long $\tau$ and build up stellar mass
slowly.  This arrangment of \taumods\ has long been indicated
by the color-magnitude relation for blue galaxies (e.g. Tully \etal 1982).
The model correctly predicts that the massive
galaxies fade substantially from $z \sim 1$ to now, 
while low mass galaxies fade little since they are still
building stellar mass.  The decrease of the global SFR since
$z\sim 1$ is thus dominated by the decrease in SFR in massive galaxies,
while the activity of low-mass galaxies changes relatively little
in the mean.  Thus the Tully-Fisher slope evolution, as differential 
luminosity evolution, is a natural outcome of an appealing model for 
star formation histories of blue galaxies.

Measurements of Tully-Fisher evolution help us to constrain
models of galaxy evolution by relating
galaxies at one epoch to another, tracking the evolutionary
descent of galaxies.  In this paper we have 
taken the simplest approach of assuming pure luminosity
evolution, i.e. that galaxies' characteristic velocities
evolve much less than their luminosities, and that the 
blue galaxy population is essentially constant in number.  
Future work on this subject can be more sophisticated.  
Theoretical ideas about the evolution of velocity with halo
mass can improve this approach.  The color-magnitude
relation, infrared magnitudes, stellar mass estimates, 
and the luminosity and velocity functions provide a wealth
of data which models must confront.  The scatter about these
relations should be related to episodic variations about the
mean star formation histories.  Together these may be used
to understand the mass assembly history of galaxies.

\acknowledgments
We dedicate this paper to the memory of Bob Schommer.
We thank the TKRS, GOODS, and Hawaii groups for making
their catalogs and data publicly available.  
We thank Greg Novak for helpful discussions on least-squares fitting.
BJW has been supported by grant NSF AST-0242860 to S. Veilleux.  
The TKRS was supported by NSF grant AST-0331730 and this 
project has been supported by NSF grant AST-0071198 to UCSC.
AJM acknowledges support from NSF grant AST-0302153 through the 
NSF Astronomy and Astrophysics Postdoctoral Fellows Program.
The authors wish to recognize and acknowledge the 
cultural role and reverence that the summit of Mauna Kea
has always had within the indigenous Hawaiian community.
We are most fortunate to have the opportunity to conduct
observations from this mountain.


%

\appendix

\section{Maximum likelihood fitting of data with intrinsic scatter}
\label{app-maxlike}

There are a number of approaches to fitting straight lines
to data with errors in both coordinates, and a perhaps
surprising lack of consensus on the ``best'' method
(see e.g. Akritas \& Bershady 1996; Gull 1989; Isobe \etal\ 1990;
Tremaine \etal\ 2002; Novak \etal\ 2006).
This problem becomes more acute when the data
have significant intrinsic scatter beyond the observational errors.  
Fitting such data with
a model that does not account for intrinsic scatter will
generally yield biased fits.  Recent approaches to
fitting with intrinsic scatter are outlined by Akritas \& Bershady (1996)
and Novak \etal\ (2006).  Here we derive a maximum likelihood
with scatter (MLS) method; a maximum likelihood
approach was also used by Willick (1999).  For similarly
motivated approaches from general Bayesian considerations,
see Reichart (2001) and d'Agostini (2005); the latter derives
a formula very similar to ours.

Suppose a model with a linear relationship in $(x,y)$ with ridgeline 
$y_{pred} = A + Bx$ and gaussian intrinsic scatter $S_y$ in the $y$-coordinate,
where the scatter is assumed to be independent of $x$.\footnote{The 
scatter could be made a function of $x$ or $y$ at the expense of 
more model parameters and a significant decrease in elegance.}
The probability density distribution of objects drawn from this 
model is:

\begin{equation}
P_{mod} (x,y) = \frac{1}{\sqrt{2\pi} S_y} 
   exp(-\frac{(y-y_{pred})^2}{2 S_y^2}) \times P_{dist}(x)
\label{eqn-probdens}
\end{equation}

\noindent
$P_{dist}(x)$ is the probability density of $x$-values of the population;
it could be be used to express the limits of the distribution 
or selection limits in $x$.  For the basic model 
without selection limits we will simplify by taking a uniform $P_{dist}(x)=1$.
Nonlinear relations $y_{pred}(x)$ can also be accomodated.

Note that, for linear $y_{pred}(x)$, 
the model distribution has a gaussian cross-section
in $x$ as well as in $y$.  In fact we could 
substitute $x_{pred} = \alpha + \beta y$,
with $\alpha = -A/B$, $\beta=1/B$, and $S_x=S_y/B$ to obtain 
the same functional form with $x$ and $y$ exchanged, and a 
prefactor of $\beta$ that 
normalizes the probability density.  Either way
one writes the distribution, it represents a set of parallel
contours of probability: a ridgeline with probability 
decreasing away from the ridge.  Thus, for a 
uniform $P_{dist}(x)$ and no selection limits, there
is nothing about this parametrization which makes a
distinction between scatter in $x$ and in $y$.  Rather,
choosing whether to assign scatter to $x$ or $y$ changes
the covariance among the intercept, slope and scatter
parameters; one computes probability in a different 3-space,
$P(A,B,S_y)$ or $P(\alpha,\beta,S_x)$.  
In practice, one should assign the scatter
to the variable which does not have strong selection limits, 
e.g. velocity in the Tully-Fisher application.

For a set of observations $x_i,y_i$, with errors $e_{xi},e_{yi}$,
each observation represents a probability density distribution
$P_{obs,i}(x,y)$.  The simplest form for the $P_{obs,i}$ is an
elliptical gaussian with independent errors,
%
%
but this formalism can accomodate covariant and non-gaussian errors.
In a Bayesian sense, the probability of the model given the data is
given by an application of Bayes's Theorem,
convolving the model distribution with the probability
of each data point and the prior probability of the parameters:

\begin{equation}
P(A,B,S_y | x_i,y_i) \propto
  P_{prior}(A,B,S_y) \times \prod_{i} ~(P_{mod} \circ P_{obs,i}),
\label{eqn-probconv}
\end{equation}

\noindent 
where 
$P_{mod} \circ P_{obs,i} = \int\int dx\ dy\ P_{mod}(x,y) P_{obs,i}(x,y)$.
If we don't have any preconceptions about the parameters we
can use a uniform $P_{prior} = 1$.  The best-fit parameters
can be found by maximizing the conditional probability
$P(A,B,S_y | x_i,y_i)$ over the parameter space of $(A,B,S_y)$.
It is convenient to work with $L(A,B,S_y) = {\rm ln}~P(A,B,S_y | x_i,y_i)$.
If (1) the model has gaussian scatter $S_y$, (2) the measurement
errors are independent in $x$ and $y$, making the
$P_{obs,i}$ elliptical gaussians, and (3) there are no 
selection effects that limit the region over which we can integrate
the convolution, $L$ has a very convenient closed form:

\begin{equation}
L = -\Sigma \frac{(y_i - (A + Bx_i))^2}{(B^2 e_{xi}^2 + e_{yi}^2 + S_y^2)} + 
{\rm constant}.
\label{eqn-maxlike}
\end{equation}

Equation \ref{eqn-maxlike} is immediately recognizable as a
generalization of the $\chi^2$ minimization
least-squares fitting formula of Press \etal\ (1992),
by adding the intrinsic scatter $S_y$ in quadrature to the $y$-error
$e_{yi}$.  This method (GLS) has been used in astronomy by 
Tremaine \etal (2002), Pizagno \etal\ (2005), and tested by 
Novak \etal (2006); these authors determine the value of $S_y$ by 
requiring that total $\chi^2/N = 1$.
The present derivation advances the GLS formula by giving it a
firmer statistical justification, making the assumptions more
transparent, and demonstrating that the choice of whether to
add the scatter in $x$ or $y$ is not {\it ad hoc,} but encodes
a choice about the covariance of the parameters.  Using the GLS
formula is effectively accepting both the assumptions made
by the MLS method and the assumptions we made above to derive
Equation \ref{eqn-maxlike}.
A Fortran program, {\tt mlsfit.f}, which performs the MLS fit
over a grid of parameters using either Equation \ref{eqn-maxlike}
or the Gaussian form of the convolution of Equation \ref{eqn-probconv},
and which can be modified for other convolutions, is available 
from the authors.

A useful property of this derivation is that integrating
the probability density convolution in Equation \ref{eqn-probconv},
rather than using Equation \ref{eqn-maxlike}, easily 
accomodates non-gaussian scatter, or arbitrary error distributions
including non-gaussian and covariant errors.  Additionally
it can be used to fit over more suitable variables; in the body 
of this paper we use the velocity dispersion (see also
Equation \ref{eqn-siginst}):

\begin{equation}
y = {\rm log}~\sigoned = {\rm log}
  \left(\frac{c}{\lambda_{obs}}\sqrt{\sigma_{obs}^2 - \sigma_{inst}^2}\right).
\end{equation}

\noindent
This equation is ill-behaved for kinematically poorly resolved
galaxies, when $\sigma_{obs}$ is close to or
less than $\sigma_{inst}$.  By changing variables and integrating
Equation \ref{eqn-probconv} over $\sigma_{obs}$, the formalism
handles these galaxies without numerical singularities, although 
the intrinsic Tully-Fisher scatter becomes non-Gaussian in
$\sigma_{obs}$ and 
the convolution integrals become computationally expensive.

We can also compute the probability $P(A,B,S_y | x_i,y_i)$ over a grid
of parameters and use this to find the {\it expectation value}
of the parameters, which is potentially more meaningful than
the location of maximum likelihood (peak conditional probability).  
However, the two are very close for well constrained models with gaussian
scatter and errors.  Computing $P(A,B,S_y | x_i,y_i)$ also allows 
us to compute confidence intervals by finding contours of
probability in the parameter space, or e.g. the interval of 
$A,B,$ or $S_y$ that contains 68\% of the probability.  These confidence
intervals allow estimates of the error on parameters $A,B,$
and $S_y$.  Using the extended least-squares method allows an
estimate of the error on scatter $S_y$ through the change in $\chi^2$ 
at the best-fit $(A,B)$ (Novak \etal\ 2006), but this method does
not generalize easily to complex contours of probability.
(However, the value of the intrinsic
scatter $S_y$ is, as always, sensitive to the accuracy of the
error estimates on individual measurements, and usually less
robust than the best fit $A$ and $B$.)

For the Tully-Fisher application in the present paper, the 
sample is subject to magnitude limits, so assumption (3) above
is violated.  This could lead to incompleteness bias, but our
simulations discussed in Section \ref{sec-fittest} show that the
effect is negligible for this sample, because the
range of the data in magnitude is much larger than the magnitude
errors.  However, the selection limits and the shallow slope
of the inverse TF relation $V(M)$ dictate that we should
add the intrinsic scatter in velocity, not in magnitude.


\begin{references}

\bigskip

{\small

\reference{} Abraham, R.G., van den Bergh, S., Glazebrook, K., 
Ellis, R.S., Santiago, B.X., Surma, P., \& Griffiths, R.E.\ 1996, \apjs, 107, 1

\reference{} Akritas, M.G., \& Bershady, M.A.\ 1996, \apj, 470, 706

\reference{} Bamford, S.P., Milvang-Jensen, B., Arag{\'o}n-Salamanca, A., 
\& Simard, L.\ 2005, \mnras, 361, 109

\reference{} Bamford, S.P., Arag{\'o}n-Salamanca, A., \&
Milvang-Jensen, B.  2006, \mnras, 366, 308

\reference{} Barton, E.J., Geller, M.J., Bromley, B.C., 
van Zee, L., \& Kenyon, S.J.\ 2001, \aj, 121, 625 


\reference{} Bauer, A.E., Drory, N., Hill, G.J., \& Feulner, G.\ 
2005, \apjl, 621, L89

\reference{} Bell, E.~F., et al.\ 2005, \apj, 625, 23 

\reference{} Bell, E. F., Wolf, C., Meisenheimer, K., Rix, H.-W., 
Borch, A., Dye, S., Kleinheinrich, M., \& McIntosh, D. H. 2004, ApJ, 608, 752



\reference{} Blanton, M.R. 2005, \apj, submitted, astro-ph/0512127

\reference{} B{\" o}hm, A., et al.\ 2004, \aap, 420, 97 

\reference{} Brinchmann, J., et al.\ 1998, \apj, 499, 112

\reference{} Brinchmann, J., Charlot, S., White, S.D.M., Tremonti, C., 
Kauffmann, G., Heckman, T., \& Brinkmann, J.\ 2004, \mnras, 351, 1151

\reference{} Broadhurst, T.J., Ellis, R.S., \& Shanks, T.\ 1988, 
\mnras, 235, 827

\reference{} Capak, P., et al.\ 2004, \aj, 127, 180 

\reference{} Conselice, C.J., Bundy, K., Ellis, R.S., Brinchmann, J., 
Vogt, N.P., \& Phillips, A.C.\ 2005, \apj, 628, 160

\reference{} Cooper, M.C. et al.\ 2006, MNRAS in press, astro-ph/0603177

\reference{} Cowie, L.L., Songaila, A., Hu, E.M., \& 
Cohen, J.G.\ 1996, \aj, 112, 839

\reference{} d'Agostini, G.  2005, preprint physics/0511182

\reference{} Dalcanton, J.J., Spergel, D.N., \& Summers, F.J.\ 
1997, \apj, 482, 659



\reference{} Faber, S.M. \etal\ 2006, \apj, submitted, astro-ph/0506044

\reference{} Fioc, M., \& Rocca-Volmerange, B.\ 1997, \aap, 326, 950 

\reference{} Flores, H., Hammer, F., Puech, M., Amram, P. \& Balkowski, C.
2006, \aap, in press, astro-ph/0603563

\reference{} Forbes, D.A., Phillips, A.C., Koo, D.C., \& 
Illingworth, G.D.\ 1996, \apj, 462, 89

\reference{} Fouque, P., Bottinelli, L., Gouguenheim, L., \& Paturel, G.\ 
1990, \apj, 349, 1 

\reference{} Giavalisco, M., et al.\ 2004, \apjl, 600, L93 

\reference{} Giovanelli, R., Haynes, M.P., Herter, T., Vogt, N.P., 
da Costa, L.N., Freudling, W., Salzer, J.J., \& Wegner, G.\ 
1997, \aj, 113, 53 
 

\reference{} Gull, S. F. 1989, in Maximum Entropy and Bayesian Methods,
ed. J. Skilling (Dordrecht: Kluwer), 511

\reference{} Isobe, T., Feigelson, E.D., Akritas, M.G., \& Babu, G.J.
1990, \apj, 364, 104

\reference{} Kannappan, S.J., \& Barton, E.J.\ 2004, \aj, 127, 2694 

\reference{} Kannappan, S.J., Fabricant, D.G., \& Franx, M.\ 
2002, \aj, 123, 2358



\reference{} Kobulnicky, H.A., \& Gebhardt, K.\ 2000, \aj, 119, 1608 

\reference{} Kobulnicky, H.A., \& Kewley, L.J.\ 2004, \apj, 617, 240

\reference{} Kobulnicky, H.A., et al.\ 2003, \apj, 599, 1006


\reference{} Larson, R.B., \& Tinsley, B.M.\ 1978, \apj, 219, 46


\reference{} Lilly, S.J., Le Fevre, O., Hammer, F., \& 
Crampton, D.\ 1996, \apjl, 460, L1

\reference{} Mall{\' e}n-Ornelas, G., Lilly, S.J., Crampton, D., 
\& Schade, D.\ 1999, \apjl, 518, L83 

\reference{} Melbourne, J., Koo, D.C., \& Le Floc'h, E.\ 2005, 
\apjl, 632, L65

\reference{} Melbourne, J., Phillips, A.C, Harker, J., Novak, G.,
Koo, D.C., \& Faber, S.M.  2006, \apj, submitted

\reference{} Metevier, A.J., Koo, D.C., Simard, L., \& Phillips, A.C.
2006, \apj, 643, 764 

\reference{} Milvang-Jensen, B., Arag{\' o}n-Salamanca, A., Hau, G.K.T., 
J{\o}rgensen, I., \& Hjorth, J.\ 2003, \mnras, 339, L1 

\reference{} Mo, H.J., Mao, S., \& White, S.D.M.\ 1998, \mnras, 295, 319

\reference{} Nakamura, O., Arag{\' o}n-Salamanca, A., Milvang-Jensen, B.,
Arimoto, N., Ikuta, C., \& Bamford, S.P. 2006, \mnras, 366, 144

\reference{} Novak, G.S., Faber, S.M., \& Dekel, A.\ 2006, \apj, 637, 96



\reference{} Pisano, D.J., Kobulnicky, H.A., Guzm{\' a}n, R., 
Gallego, J., \& Bershady, M.A.\ 2001, \aj, 122, 1194 

\reference{} Pizagno, J. \etal\  2005, \apj, 633, 844

\reference{} Portinari, L., \& Sommer-Larsen, J.  2006, astro-ph/0606531

\reference{} Press, W.H., Flannery, B.P., Teukolsky, S.A. \& Vetterling,
W.T. 1992, Numerical Recipes,  (Cambridge: Cambridge U.P.)

\reference{} Reichart, D.E.\ 2001, \apj, 553, 235 

\reference{} Rix, H., Guhathakurta, P., 
Colless, M., \& Ing, K.\ 1997, \mnras, 285, 779 

\reference{} Sakai, S. et al.\ 2000, \apj, 529, 698


\reference{} Schechter, P.L.\ 1980, \aj, 85, 801 

\reference{} Searle, L., Sargent, W.L.W., \& Bagnuolo, W.G.\ 
1973, \apj, 179, 427 

\reference{} Simard, L., \& Pritchet, C.J.\ 1998, \apj, 505, 96 


\reference{} Steinmetz, M., \& Navarro, J.F.\ 1999, \apj, 513, 555 


\reference{} Teerikorpi, P.\ 1987, \aap, 173, 39 

\reference{} Tremaine, S., et al.\ 2002, \apj, 574, 740 


\reference{} Tully, R.B., Mould, J.R., \& Aaronson, M.\ 1982, \apj, 257, 527

\reference{} Tully, R.B. \& Pierce, M.J.  2000, \apj, 533, 744

\reference{} Tully, R.B., Pierce, M.J., Huang, J., Saunders, W., 
Verheijen, M.A.W., \& Witchalls, P.L.\ 1998, \aj, 115, 2264

\reference{} Vogt, N.P., et al.\ 1996, \apj, 465, L15

\reference{} Vogt, N.P., et al.\ 1997, \apj, 479, L121

\reference{} Vogt, N.P.\ 2000, ASP Conf.~Ser.~197: Dynamics of 
Galaxies: from the Early Universe to the Present, 197, 435

\reference{} Watanabe, M., Yasuda, N., Itoh, N., 
Ichikawa, T., \& Yanagisawa, K.\ 2001, \apj, 555, 215 

\reference{} Wechsler, R.H., Bullock, J.S., Primack, J.R., Kravtsov, A.V., 
\& Dekel, A.\ 2002, \apj, 568, 52 

\reference{} Weiner, B.J., et al.\ 2005, \apj, 620, 595

\reference{} Weiner, B.J., et al.\ 2006, \apj, (Paper I)

\reference{} Willick, J.A.\ 1994, \apjs, 92, 1

\reference{} Willick, J.A.\ 1999, \apj, 516, 47

\reference{} Willmer, C.N.A., et al.\ 2006, \apj, in press

\reference{} Wirth, G.D., et al.\ 2004, \aj, 127, 3121

\reference{} Wolf, C., Meisenheimer, K., Rix, H.-W., Borch, A., 
Dye, S., \& Kleinheinrich, M.\ 2003, \aap, 401, 73

\reference{} Ziegler, B.L., et al.\ 2002, \apjl, 564, L69 

}

\end{references}
\end{document}